\documentclass[showpacs,preprintnumbers,amsmath,amssymb]{revtex4}
\usepackage{graphicx}
\usepackage{amsfonts}
\catcode`\Ž=\active \def Ž{\'e}

\begin{document}
\title{Black Hole Induced Ejections}
\author{Guy Pelletier}
\affiliation{Laboratoire d'Astrophysique de Grenoble, and Institut Universitaire de France
}
\date{March 2004}

\begin{abstract}
    Black Holes generate a particular kind of environment dominated 
    by an accretion flow which concentrates a magnetic field. The 
    interplay of gravity and magnetism creates this paradoxical 
    situation where relativistic ejection is allowed and consequently high 
    energy phenomena take place. Therefore Black Holes, which are very likely 
    at the origin of powerfull astrophysical phenomena  such as AGNs, micro-
    quasars and GRBs where relativistic ejections are observed, are at the heart 
    of high energy astrophysics. The 
    combination of General Relativity and Magneto-HydroDynamics (MHD) 
    makes theory difficult; however great pionneers opened beautiful 
    tracks in the seventies and left important problems to be solved 
    for the next decades. These lectures will present the status of 
    these issues. They have a tutorial aspect together with a critical review aspect 
    and contain also some new issues.
    Most of these lectures has been presented at the school on "Black Holes in the Universe" 
   at Carg\`ese, in May 2003.
\end{abstract}

\maketitle

\section{Introduction}\label{sec.INTRO}

The existence of Black Holes is now very likely and established for some interesting cases: at the Galactic Centre, where not only the mass is known but also the spin; in the nucleus of some Seyfert galaxies where either maser effects have been observed in the accretion disk or where iron lines have been analysed and display a strong Doppler effect, that reveals a large mass concentration. The physics of Black Hole environments is getting more and more grounded and receives constant progress thanks to new generation of instruments, especially in X-ray and Gamma-ray astronomy.

The generation of relativistic jets is probably one of the most fascinating phenomenon attributed to Black Hole environments, where the accretion process governs most of the phenomenology. The fact that the most energetic phenomena in the Universe come from plasma ejections in the vicinity of Black Holes is a strong paradox that stimulates challenging theoretical investigations.

Moreover, these relativistic ejections generate high energy phenomena, that are at the forefront of the astroparticles physics, which motivates the development of a new generation of instruments that flourish everywhere in the World.

These lectures will introduce all the physics useful to built an understanding of what is happening with the relativistic jets in Active Galactic Nuclei (hereafter AGNs) and micro-quasar jets, as well as in Gamma-Ray Bursts (hereafter GRBs). These topics incite the development of the physics of relativistic plasmas and relativistic MHD, which are one of the major front of modern plasma physics.

These astrophysics topics are recurrently reviewed (\cite{BBR} \cite{CMa} \cite{CMb}). However all the useful background of knowledge is briefly recalled, in order that the presentation be self-contained and read by a beginner. Furthermore, more detailed developments will be oriented towards numerous references, and some opened questions and trends will be presented. The content of the chapter is organised as follows. The main observational features of the relativistic ejections (AGNs, microQuasars and GRBs) are reviewed in section (2),  with a presentation of the order of magnitudes of the main parameters, and of their performances as high energy engines. The physical properties of spinning Black Holes and the electromagnetism in Kerr metric is briefly recalled in section (3) in order to prepare the reader to the physics of jet generation. More detailed presentations can be found in text books (see for instance \cite{MTW}, \cite{ZN}, \cite{TPD}). Section (4) is devoted to the theoretical description of jets, including the presentation of the Blandford-Znajek mechanism of jet generation from a Kerr Black Hole. The properties of relativistic jets are different from those of non-relativistic jets, and they are emphasised in section (5), notably the difference in their confinement properties, the issue of the Compton drag suffered by relativistic jets, and the issue of the baryonic pollution of a pair jet. The other mechanism of jet generation from an accretion disk is presented in section (6), with an emphasis on the conditions, especially on the transport coefficients, to get launching.
A presentation of the "two-flow" model is given in section (7) which consists in combining both a mildly or subrelativistic jet with relativistic ejections in the core, the subrelativistic jet playing a crucial role to circumvent the main difficulties encountered in the theoretical attempt to describe the generation of relativistic jets. The status of this field of research and of its new trends are summarised as a conclusion.

\section{Relativistic Ejections}\label{sec.RELJET}

\subsection{AGN Jets}\label{subsec.AGNJ}

Some Active Galactic Nuclei (AGNs) exhibits powerful collimated jets. Although very menority, they are very interesting because, first the ejection phenomenon from a Black Hole environment is very intriguing, and also because jets are sources of high energy processes. There are two classes of jets, namely FR1 and FR2 jets (the classification is due to Fanarov -Riley (see \cite{BBR})). FR2 jets are the most powerful, very collimated and ending with hot spots, whereas FR1 are weaker, shorter, displaying wiggles and no hot spot.

From radio observations and Chandra measurements, the parameters of some FR2 jet hot spots are rather well estimated, such as Cygnus A and Pictor A hot spots (Wilson et al. \cite{WIL}). Consider Cygnus A; the size ($R_{HS} \simeq 3\, kpc$) and the advance speed ($V_{HS} \simeq 0.1c$) are known for a long time, the radio synchrotron emission has been mapped by VLA with accurate isophotes and the synchrotron self-Compton emission (SSC) is now measured in X-rays by Chandra. Thus the magnetic field intensity B can be estimated by the ratio of the luminosities, $L_{SSC}$ over $L_{syn}$; and $B \simeq 2\times 10^{-4} G$, a value close to the equipartition estimate. Assuming that the hot spots are powered by a mildly relativistic jet ($\beta_jc$), the jet power can be estimated and 
\begin{equation}
\label{PJ}
P_j \simeq 6 \beta_j \times 10^{45} erg/s \ .
\end{equation}
This number is very comparable with the accretion power around a Black Hole of a few $10^8 M_{\odot}$ and thus indicates that a sizable fraction of the accretion power actually goes into the jets generation. 

In  fact, as will be analysed later on in section (6), when an accretion disk launches jets, a sizeable fraction of the accretion power goes in the jets, and the bulk velocity essentially depends on the fraction of mass flux that is rerouted from the accretion flow to the outflow.

Some population of AGNs exhibits high energy phenomena, such as hard X-ray emission and Gamma-ray emission, sometimes up to TeV range, together with a fast variability. These emissions come from a relativistic jet pointing towards the observer, as will be detailed in the subsection (2.3). Those sources are called "Blazars", which includes two categories of objects: flat spectrum radio quasars (FSRQ) and BL-Lacs, themselves divided in high BL-Lacs and low BL-Lacs.

\subsection{Eddington scaling}\label{subsec.EDDSC}

Given the mass M of the black hole, several units of this physics can be defined (\cite{BBR}). The gravitational radius $r_{g} \equiv GM/c^{2}$ is the basic length unit, from which an obvious time scale stems, namely $\tau_{g} \equiv r_{g}/c$. The last stable orbit is at $6r_{g}$ for a 
Schwarzschild Black Hole, at $r_{g}$ for a Kerr Black Hole at maximum rotation. The environment is made of a hot plasma undergoing an accretion flow and sometimes an ejection flow, both manisfesting through their radiation. The powers associated with these flows and radiation tightly depend on the accretion rate $\dot M_{a}$. Through the accretion flow, matter progressively
goes deeper and deeper in an effective potential well that describes the effects of both gravitation and  rotation. In Newtonian gravity, this is simply $\Phi_{eff}(r) = -GM/r + l^2/2r^2$ where $l$ is the specific angular momentum of the matter element. The effective potential is deeply modified by General Relativity effects, described by the two Black Hole parameters, its mass $M$ and its reduced spin parameter $a_s$ ($\vert a_s\vert \leq 1$). Accretion power characterises the rate of energy liberated by matter flowing from infinity down to the last stable circular orbit; below the last stable orbit, matter falls almost freely towards the Black Hole horizon, without significantly radiating. In Newtonian dynamics, the accretion power is
\begin{equation}
    P_{a} = \frac{1}{2}\dot M_{a}\frac{GM}{r_{lso}} \ .
      \label{eq:PACCR}
\end{equation}
More generally, it is given by
\begin{equation}
\label{PACM }
 P_{a} =   \varepsilon_{a} \dot M_{a}c^{2} \ ,
\end{equation}
where $\varepsilon_a$ is called the accretion efficiency and includes relativiqst¤ic effects; its value is $6 \%$ for a Schwarzschild Black Hole, $42 \%$ for a Kerr Black at maximum rotation.

The Eddington luminosity is a useful reference for the accretion power:
\begin{equation}
    L \leq L_{E} = \frac{4\pi c}{\sigma_{T}} GMm_{p} \simeq 1.3 \times 10^{46} 
    (\frac{M}{10^{8}M_{\odot}}) ergs/s
    \label{eq:LUMED}
\end{equation}
For a given efficiency, it defines also a reference for the accretion 
rate: $\dot M_{a} \equiv \dot m \dot M_{Ed}$ and $P_{a} = \dot m L_{Ed}$.

The accretion power is shared between the radiation flux of the disk 
and the power in the jets it launches.
$L = \chi P_{a}$ and $P_{j} = {1\over 2}(1-\chi) \, P_{a}$.

The luminosity emitted by the disk depends on its opacity, the most powerful emission occurs when the disk is optically thick. In the hottest part of the disk and of the jet, the Compton opacity is the relevant one: $\tau = \sigma_{T}n r_0$. In standard accretion disk (hereafter SAD) $\tau>1$ (\cite{SS}), whereas weakly radiative tori with $\tau<1$ has also been considered, as, for instance, advective dominated accretion flow (hereafter ADAF) (\cite{NY}).

The black body temperature of the SAD is maximum at a few last 
stable orbit radius, $8 r_{g}$ typically (for $a_s=0$) and determines the typical 
frequency range of thermal emission.
\begin{equation}
    T_{*} \simeq \frac{1}{4\times 2^{1/4}}(\frac{L}{\pi \sigma_{st} 
    r_{g}^{2}})^{1/4} \ .
    \label{eq:TMAX}
\end{equation}
For the Eddington luminosity, 
\begin{equation}
    T_{*} \simeq 10 \times (\frac{M}{10^{8}M_{\odot}})^{1/4} eV \ .
    \label{eq:TMAX}
\end{equation}
The corresponding maximum pressure is $P_{*} = 
\frac{1}{3}a_{st}T_{*}^{4} \simeq 3\times 10^{-2} L/(\pi r_{g}^{2}c)$.

The magnetic field plays a key role together with gravitation in the physics of accretion, ejection and its associated high energy manifestation. It is concentrated by the accretion flow; Dynamo action could also be at work in accretion disk (\cite{CMa}). For physical reasons, it is convenient to compare its intensity with
the value obtained by assuming equipartition of its pressure with the kinetic pressure (of the plasma or the photons) on the equatorial plane of the accretion flow (an excess of magnetic field close to the central black hole would trigger interchange instabilities that would expel the excess).
For a Schwarzschild Black Hole, the equipartition magnetic field for Eddington 
luminosity is thus
\begin{equation}
    B_{*} \simeq 1.2 \times 10^{4} (\frac{M}{10^{8}M_{\odot}})^{-1/2}  G \ .
    \label{eq:BEDD}
\end{equation}
In the case of optically thin tori (which is likely the accretion regime close to a spinning black hole \cite{ACGL} \cite{CAL}), the magnetic field pressure does not exceed a fraction of $\rho \Omega^2 r^2$, and, because the mass density $\rho$ is much weaker than in the case of an optically thick disk ($n\sigma_T h \ll 1$ with $h \sim r$), the magnetic field intensity can be stronger than the previous estimate, but not more than a factor ten, and follows the same scaling with $M$ (however its dependence with $a_s$ need some further examination). The electric current flowing in the accretion-ejection system is particular important as will be seen further on; ($B_*r_g$) can be considered as its fiducial value. The magnetic field and the current together  with rotation allow to develop a power for the generation of jets. The magnetic field plays an important role in the transport phenomena (diffusion, effective viscosity, thermal conduction etc.) and in the processes of particle acceleration. With simple arguments, a fiducial number can be proposed for the maximum energy of the Cosmic Rays accelerated in the jets. MHD perturbations (waves, reconnections and shocks) in the jets can no more accelerate particles whose Larmor radii would exceed the radius of the jets; the following maximum energy of Cosmic Rays is thus obtained:
$\epsilon_{max} \simeq 10^{20} \Gamma Z(BR)\, eV$ (with $B$ measured in Gauss and $R$ in pc); and 
$BR_{j} \sim (\frac{M}{10^{8}M_{\odot}})^{1/2} G\times pc$
$$
\epsilon_{max} \simeq 10^{20} \Gamma Z(\frac{M}{10^{8}M_{\odot}})^{1/2} eV
$$
As a result of particle acceleration, all kinds of high energy radiations could be emitted by the jets: synchrotron emission by relativistic electrons from radio band up to IR, UV, and sometimes X-rays ranges, inverse Compton emission in gamma ray range, sometimes up to TeV range, p$\gamma$-Neutrinos emission as a signature of very high energy cosmic rays. Black Hole formation and jet acceleration generate gravitational waves that could hopefully be detected in the future (\cite{VP}).

These sources of radiations are compact in a precise physical sense, and characterised by the "compactness parameter":
\begin{equation}
    \ell_{c}(r_{0}) \equiv \frac{\sigma_{T}L}{4\pi m_{e}c^{3}r_{0}} = 
    \frac{m_{p}}{m_{e}}\frac{L}{L_{Ed}}\frac{r_{g}}{r_0} \ .
\end{equation}
The compactness is related to the opacity to $\gamma$-rays due to the process of pair creation by photon-photon interaction. The threshold of the 
process is such that $\epsilon_1 \epsilon_2 > 2/(1-\cos \theta)$ where $\epsilon \equiv h\nu/m_{e}c^{2}$. 
A radiation field of soft photons, of energies $\epsilon_s$, with a spectrum $S(\epsilon_{s}) \propto \epsilon_{s}^{-\alpha}$, constitutes an absorbing medium for gamma photon.  For gamma-photons of energies $\epsilon$, the opacity to pair creation is given by $\tau_{\gamma \gamma} \simeq 0.2 \ell_{c} \epsilon^{\alpha}$ (\cite{S}). Thus the higher the energy of the $\gamma$-photon the stronger the absorption.
The nuclei of AGN and of microquasars are clearly optically thick to gamma rays since the compactness is $10^2-10^3$ and can be richly populated with $e^+e^-$-pairs.
Compactness characterises not only the opacity to gamma-rays, but also measures the strength of the Compton drag exerted on a relativistic flow, as will be seen later on in subsection (5.1) and section (7).

\subsection{Gamma-Ray Emission and Relativistic Kinematics}

Gamma rays can escape only at remote distance and thanks to relativistic motions.
A source of photons in relativistic motions ($\beta_0$ close to 1) with a bulk Lorentz factor $\Gamma \equiv (1-\beta_0^2)^{-1/2} > 2$, is a cause of a Doppler shift, of a Doppler beaming, of  an apparent superluminal velocity and a shorter apparent variability. 
\begin{itemize}
  \item 
The photons energy are blue shifted when the flow points toward the observer at an angle $\theta$:
\begin{equation}
    \nu'=\Gamma (1-\beta \cos \theta)\nu \equiv \nu/\delta \ .
    \label{eq:EFDOP}
\end{equation}
Synchrotron and Compton peaks are blue shifted $\nu_{obs} = 
\delta\nu_{com}$.
  \item 
The variability time scale appears much shorter than 
in the co-moving flow: $\Delta t_{obs} = \Delta t_{com}/\delta$.
  \item 
The emission is beamed. Because $I(\nu, \vec n)/\nu^{3}$ is proportional to the number of 
photons in the state defined by $\nu$ and $\vec n$, it is a Lorentz 
invariant. Thus 
\begin{equation}
    I(\nu, \vec n) =\delta^{3} I_{com}(\nu/\delta) \ .
    \label{eq:DOPBEA}
\end{equation}
\item
The velocity of the source can appear to be faster than the speed of light. Indeed, during $\delta t$, the source travels a distance $v \delta t \sin \theta$ as projected on the sky plane; whereas the apparent duration of the travel is $\delta t_{app} = (1-\beta_0 \cos \theta)\delta t$. The apparent velocity is thus 
\begin{equation}
\label{VAPP}
v_{app} = v \frac{\sin \theta}{1-\beta_0 \cos \theta} \ , 
\end{equation}
and reaches a maximum $\Gamma v$ for $\cos \theta = \beta_0$ or $\sin \theta = 1/\Gamma$.
\end{itemize}

Moreover, if the source of observed photons is moving relativistically, 
the estimate of the intrinsic compactness must be rescaled compared to 
the isotropic assumption made for calculating the luminosity 
($L'=\delta^{-4}L$, $E'= \delta E$, $\Delta t'=\delta \Delta t_{obs}$) 
so that $\ell_{c} = \delta^{-5} \ell_{c}\vert_{obs}$. For $\alpha = 
1$, $\tau_{\gamma \gamma} = \delta^{-6} \tau_{\gamma \gamma}\vert_{obs}$.
For AGN, observation of GeV photons requires $\delta > 5$. For GRBs, the compactness is very high ($\ell_c \sim 10^{12 }$) and
observation of $10\, MeV$ photons requires $\delta >100$. The opacity to pair creation leads to an observed break in gamma-spectra in MeV range, at
$$
\epsilon_{b} = \delta  (0.2\, \ell_{c})^{-1/\alpha} \ .
$$
These relativity kinematics effects are crucial to analyse the Blazar spectra with their double peaks, synchrotron peak in X-rays and SSC peak in gamma-ray, together with variability (see \cite{KKS} for recent fits).
The observation of gamma rays is a strong argument in favour of relativistic jets.
This argument adds to the historical argument inferred by Martin Rees in 1966 (\cite{R}) in favour of relativistic motions in AGNs in order to prevent the Compton catastrophe (\cite{BBR}).

\subsection{Microquasars}\label{subsec.MICROQ}

Galactic sources, which have been discovered during the last decade \cite{MR1} \cite{MR2}, have features similar to radio quasars : a bright nucleus with twin radio jets displaying superluminal motions and hard X-ray IC-emission; they have been called micro-quasars. 

Most of the phenomenology can be understood by simply rescaling the Eddington fiducial numbers previously introduced that depends on the Black Hole mass only: $10^7-10^9 M_{\odot}$ for AGNs and
 $\sim 10 M_{\odot}$ for micro-quasars. The inverse Compton process on accretion disk photons sensitively changes in the case of micro-quasars because the disk black body radiation is in the soft X-rays and thus the inverse Compton process is deeply modified by the Klein-Nishina regime (see subsection (7.2)).

At first look, the proximity of micro-quasars seems interesting, letting think that they will be studied more easily than remote AGNs. In fact, in terms of gravitational radii, the angular resolution is much better with AGNs at a few Gpc than with the closer micro-quasars. The advantage of the smallness and closeness of  micro-quasars is the much shorter time evolution. Indeed, connection between the accretion process and the ejection has been nicely revealed by the changes of regime (\cite{CM})(\cite{BEL}). These variations are very long compared to the Keplerian time at the inner orbits of the accretion disk, and similar variations could not be observed in AGNs during a human life. On the other hand, the shortest variability at Keplerian time is better recordered in AGNs, especially in Gamma-Rays; next X-ray satellites (as for instance SIMBOL-X) will probably be able to follow millisecond variations in micro-quasars.

As source of cosmic rays of energies up to $10^{16}eV$, a neutrinos emission resulting from $p\gamma$-collisions could be detected (\cite{WAX}).

\subsection{Gamma Ray Bursts}\label{subsec.GRBs}

Two kinds of progenitors have been considered to give rise to the GRB phenomena: coalescence of compact objects and collapse of supermassive stars. Besides two kinds of origination, 
there are two classes of models for the description of the GRB phenomenon itself : the "fireball" model (\cite{RM}) and the electromagnetic model (\cite{BL}). 

The fireball model is based on the assumption of an initial state of high relativistic enthalpy that gives rise to a relativistic flow through adiabatic expansion.
The energy involved in the explosion is $E \sim 10^{51} erg$; the initial temperature is therefore $T_0 \sim 5 MeV$. The dynamics depends on the baryon load of the relativistic wind; it is measured by the parameter
$\eta \equiv E/M_bc^2$. The bulk Lorentz factor saturates at the value $\eta$ after a travel from few gravitational radii ($r_0$) to $r_s = \eta r_0$. Then the ejected shells start broadening at $r_b = \eta^2 r_0$. Nowadays, the value $\eta \sim 300$ allows the best fit of most of the observed GRBs, but no physical argument to select this value has been proposed so far.

To explain the fast irregularities of the light curves of many GRBs, the fire ball model has been remodeled by considering multiple shells instead of a single one \cite{MRs}. The date of collision between two shells launched at a time interval $\Delta t$ is given by
\begin{equation}
\label{TC}
t_c = 2\frac{\Gamma_1^2 \Gamma_2^2}{\Gamma_2^2 - \Gamma_1^2} \Delta t \ .
\end{equation}

The maximum duration of the prompt emission is $\Delta t_{max} \sim 2\Gamma^2 \Delta t_w$, where $\Delta t_w$ is the duration of the relativistic ejection.
The observed duration of the prompt emission is $ \Delta t_{obs} = (1-\beta) \Delta t_{max} \simeq \Delta t_{max}/2\Gamma^2 \sim \Delta t_w$; therefore the observations give access to the duration of the ejection, and two categories of GRBs have been observed: short ones with $<\Delta t_w> \simeq 0.1\, s$ and long ones with $<\Delta t_w> \simeq 10 \, s$. Similarly, the observed variability time corresponds to irregularities at the origin of the flow with a similar time scale, $\Delta t_{obs} 
\sim \Delta t_{var}$. Millisecond variations are sometimes observed; they correspond to shell collisions at a few $r_b = \eta^2 r_0$. The proper length of the flow is $l_0 = \beta c \Gamma \Delta t_w$ which indicates that the jet length is smaller than the distance where the shells start to collide each other ($\beta c \Delta t_w < r_b$). 
The flow and the magnetic field are likely disconnected from the central engine when the prompt emission starts. Fairly collimated (for a few GRBs, the average solid angle is estimated as $\Omega/4\pi \simeq 2\times 10^{-3}$), these relativistic flow can be considered as transient jets. However there is no physical argument either for explaining the duration of the flow. They end in the form of a strong relativistic shock with the interstellar medium, or the ejecta medium, that produces the so-called "afterglow", which has been well studied and brought support to the model.

The equipartition magnetic field can be as high as $B \sim 10^{15}G$ at the initial state. The magnetic pressure decreases in the expansion like the kinetic pressure, namely $P_m \propto P \propto r^{-4}$, as long as the magnetic field is considered to be tangled, carried by the shells and disconnected from the source (\cite{MRL}). In the canonical fireball model, the magnetic field can never dominate.

In the electromagnetic model, a "magnetar" object is initially created (\cite{LB}), whose environment contains a rather cool plasma highly dominated by the magnetic field, and which is in fast rotation. Therefore the relativistic wind is no more due to adiabatic expansion, but to a powerful Poynting flux. This model is presented in another chapter of this book (see Lyutikov's and Blandford's chapters). 

GRBs are popular events for high energy astrophysics because, not only they are very powerful gamma-ray emitters, but also they are the most promising sources of UHECRs, with an associated high energy Neutrinos emission (\cite{WAX}), and probably detectable sources of Gravitational Waves (\cite{VP}).

\subsection{Performances of the Ejections}\label{subsec.PERF}
In the table \ref{perfor}, the order of magnitude of the main parameters of the three kinds of relativistic sources are indicated: their power, the mass of the central black hole, the gravitational time that indicates the fastest variability time scale, the bulk Lorentz factor of the jet, and the equipartition value of the magnetic field in the vicinity of the black hole.
\begin{table*}[h!t]     
\begin{center}     
\begin{tabular}{|c|c|c|c|c|c|}      
\hline     
 & Power (erg/s)& $M_{BH}$ ($M_{\odot}$) & $\tau_g$ (s) & $\Gamma$ & B (G) \\      
\hline     
AGNs & $ 10^{44-46}$ & $10^{7-9}$  & $10^{2-4}$& $\sim 10$ & $\sim 10^4$ \\     
\hline     
$\mu$-quasars & $\sim 10^{38}$& $\sim 10$& $10^{-4}$& $\sim 10$ & $\sim 10^{8}$ \\     
\hline     
GRBs & $10^{49-52}$& 10-100 & $\sim 10^{-3}$ & $\sim 300$ & $\sim 10^{15}$ \\     
\hline     
\end{tabular}     
\caption{Performance of the ejections.}    
\label{perfor}     
\end{center}      
\end{table*}     
Thanks to magnetic irregularities and fast motions, both processes of Fermi acceleration,  by MHD turbulence and shocks, generate ultrarelativistic electrons that radiate synchrotron emission. Lorentz factors achieved by the electrons in the co-moving frame are on the order of $10^6$ in AGN and micro-quasar jets. Therefore jets produce a synchrotron emission in Radio band and sometimes up to X-rays. The main gamma-ray emission of GRBs is produced by ultrarelativistic electrons that radiate synchrotron keV-photons in the co-moving frame.

Acceleration of cosmic ray protons is expected in these three kinds of relativistic flows. As previously explained, their energy is necessarily smaller than the confinement limit which can reach a few $10^{20}eV$ at the beginning of the AGN jets where the electric current is maximum. This concerns FR2 jets, for FR1 jets like Centaurus A do not have a sufficient product $BR_j$ (see Morganti et al. \cite{MOR}), thus the current is too low for achieving ultra high energies. The same argument for micro-quasars leads to $10^{16}eV$. AGN hot spots are also considered as possible sources of UHECRs; however this issue will not be discussed in the chapter.

In GRBs, the limiting size for the maximum cosmic ray energy is the width of the shells at the beginning of the stage of internal shocks, namely at $r_b$, where their width is $\Delta R \sim r/\Gamma$ in the co-moving frame. Thus $\epsilon_{max} \sim 10^{20} Z(Br)_b$ and since $r_b \sim 10^{12}cm$, the generation of UHECRs would require a magnetic field of order $10^6G$, and thus $\sim 10^{15}G$ at the initial state.
It turns out that the terminal shock is not able to generate the expected flux of UHECRs, but this will not be discussed in this chapter either.

\section{Spinning Black Holes and Relativistic MHD}

The main point of the jet formation is that opened magnetic field 
lines thread either an accretion disk (or torus) or a spinning black hole and the 
rotation of the field lines differs from the rotation of the matter or 
the black hole. Thus a poloidal current is driven whose associated 
toroidal field $B_{\phi}$ is responsible for 
\begin{itemize}
  \item transfering the angular momentum from the rotator through the 
magnetic stress (and torque) $-B_{\phi}B_{p}/4\pi$;
  \item generating of poloidal Poynting flux $E \times B_{\phi}/4\pi$, the 
electric field being generated by the rotation of the field lines; a 
poloidal force also pushes matter along the poloidal direction;
$\propto -\nabla (rB_{\phi})^2$
  \item a confinement is possible because of the magnetic tension 
$-B_{\phi}^{2}/4\pi$.
\end{itemize}

The general pattern of the accretion-ejection flow is sketched on figure (1). The poloidal magnetic 
field (blue lines) has a bipolar configuration (a quadrupolar configuration is inefficient to extract the 
angular momentum). The poloidal current (yellow lines), associated with the toroidal field generated by differential rotation, flows towards the central object and the inner part of the accretion disk; the current closure is generally insured at remote distance from axis with current lines emerging from the disk and unfolding in the jets (see \cite{F}).
\begin{figure}[h]
\begin{center}
\caption{Sketch of the general pattern of the MHD flow. In blue the field lines, in yellow the current, in red the electric field, in green the matter flow. The dashed line suggests the "watershed" surface.}
\label{fig1}
\end{center}    
\end{figure}
Note that the electric field (red lines) points towards the axis almost everywhere except close to the Black Hole horizon as will explained later on. Part of the accretion flow (green lines) is rerouted by the magnetic field and supplies the jet, it crosses the magnetic surfaces in the accretion dissipative region and then the outflow runs along the poloidal field lines in ideal MHD regime. Thus there is a "watershed" surface in the disk that separates the matter flow that will actually be accreted by the black hole and the matter outflow. Above the black hole, no matter would be expected; however pair creation takes place there and the watershed surface separates the pair flow that falls onto the black hole and the pair flow that is expelled along the axis.

\subsection{Summary of the "3+1" split of the Kerr metric}\label{subsec.SPLIT}

The Kerr metrics of space-time is the solution of General Relativisty equations (see \cite{MTW})
with two symmetries: stationarity and axial symmetry. Whereas the Schwarzschild
solution has a point singularity at $r=0$, the Kerr solution has a 
ring singularity at $r=0$ and $\theta = \pi/2$.
It depends on two parameters only: the mass $M$ (or $r_{g} \equiv 
GM/c^2$) and the spin parameter $a_{s}$. The 
angular momentum of the Black Hole is $J = a_{s}r_{g}Mc$. In astrophysics, the third parameter 
of black hole solutions, namely the electric charge, is always zero, in practice. The Boyer-Lindquist 
system of coordinates for a spinning black hole 
is a suitable universal frame for the space-time 4-manifold ${\cal E}_{4}$.
\begin{equation}
    ds^{2} = -\alpha^{2}dt^{2} +g_{ij}(dx^i+\beta^idt)(dx^j+\beta^jdt) \ ,
   \label{eq:BLCOOR}
\end{equation}
where $dt$ stands for $cdt$.
The main relativistic effects are described by the two following 
functions of $r$ and $\theta$: the "lapse" function $\alpha = 
h_{1}h_{2}/h_{3}$ (multiplied by $c^{2}$ in usual units) and the GM-potential 
$\omega = (2a_{s}/h_{3}^2) (r/r_{g})$ (multiplied by $c/r_g$ in usual units). The 
"frame drag", specific to Black Hole rotation, is characterised by 
$\beta^{\phi}= -\omega$ ($\beta^r = \beta^{\theta}=0$).
The metric elements are 
$$
g_{rr}=(h_{2}/h_{1})^{2} \ ,\hspace{3pt} g_{\theta \theta} = 
r_{g}^{2}h_{2}^2 \ , \hspace{3pt} g_{\phi \phi} \equiv \bar 
r^{2}=r_{g}^{2}(h_{3}/h_{2})^2\sin^{2}\theta \ ,
$$
where the three 
functions $h_{1}$, $h_{2}$ and $h_{3}$ depend on $r/r_{g}$, $\theta$ 
and $a_{s}$.
$$
h_{1}^{2}= (\frac{r}{r_{g}})^2 -2\frac{r}{r_{g}}+a_{s}^{2}
$$
$$
h_{2}^{2}= (\frac{r}{r_{g}})^2 +a_{s}^{2}\cos^{2}\theta
$$
$$
h_{3}^{2}= (\frac{r^{2}}{r_{g}^{2}} +a_{s}^2)^2 
-a_{s}^{2}h_{1}^{2}\sin^{2}\theta
$$

The BL-system of coordinates displays a universal time $t$ that allows 
to slice the four 
dimensional space-time into 3D-subspaces at constant $t$, these 
subspaces being isomorph to the Riemanian 3-manifold ${\cal F}_{3}$ 
with metric $dl^{2} = g_{ij}dx^i dx^j$, endowed with spatial curvature 
that characterises gravitation and that prevents the extension of any local 
inertial system of coordinates to a global one. At any point of ${\cal 
F}_{3}$, a "Fiducial Observer" (FIDO) is at rest with respect of the 
local time $\tau$ and thus moves with respect to the BL-system; it 
rotates at the sheared angular velocity $\omega$. The conversion of the local 
time $\tau$ such that $ds^{2}= d\tau^{2} -dl^{2}$ to the universal one 
is characterised by the lapse function 
$\alpha= d\tau/dt$ that measures the gravitational red-shift. The temporal 
curvature makes the global extension 
of the local time impossible. Note that $g_{tt} = -\alpha^2 + 
\omega^2 \bar r^{2} = -(h_{1}^{2}-a_{s}^{2} \sin^{2} \theta)/h_{2}^{2}$ 
and $g_{t\phi} = g_{\phi t} = -\omega \bar r^{2}$.

The length of the circumference defined by the set of points of given 
$(r,\theta)$ is $2\pi \bar r$; and $\bar r$ is called the "circumferencial" radius 
and reduces to $r \sin \theta$ at remote distance from the Black Hole.
The horizon of the black hole is such that $\alpha(r, \theta) = 0$; 
this is a 2-sphere of radius $r_{H} = r_{g}(1 + \sqrt{1-a_{s}^{2}})$. The area
of the horizon is $A = 4\pi (r_H^2 + a_s^2 r_g^2)$.

If one looks for FIDO that would be static, namely, at rest with 
respect to the BL-system, then, because of Black Hole rotation, it 
turns out that none exists inside the so-called "ergosphere", a 2-manifold 
such that $r= r_{g}(1+\sqrt{1-a_{s}^2\cos^{2} \theta})$ (flattened by 
rotation effect), which corresponds to $g_{tt} = 0$. They are 
constrained to rotate with an angular velocity between $\Omega_{1}$ 
and $\Omega_{2}$ where $\Omega_{1,2}= \omega \pm \alpha/\bar r$. It is 
remarkable that these two angular velocities merge at the horizon: 
$\Omega_{1}= \Omega_{2} = \omega\mid_{H}$. That angular velocity 
$\omega\mid_{H}$ is uniform over the horizon ("solid rotation") and 
can be defined as the Black Hole angular velocity $\Omega_{H}$; its 
relation to the spin parameter is given by $\Omega_{H}=a_{s}c/2r_{H}$.
FIDOs inside 
the ergosphere necessarly rotate and for the angular velocity 
$\omega$ they have no angular momentum with respect to the BL-system; 
so they are also called ZAMOs (Zero Angular Momentum Observers).

Because of stationarity with respect to the 
universal time, an energy invariant $\epsilon$ exists and because of 
axisymmetry, an angular momentum invariant $L_{z}$ exists for geodesic 
motions. In terms of Killing generators of the symmetries $V^{\mu}$, 
the invariants are $p_{\mu}V^{\mu}$. With respect to the BL-reference 
frame, 
$$
\epsilon = -p_{0} = -(g_{tt}p^{0}+g_{t\phi}p^{\phi})
$$
and
$$
L_{z} = p_{\phi} = g_{\phi \phi} p^{\phi}+g_{\phi t} p^{0}
$$
Thus, combining both relations, the energy can be written:
$$
\epsilon = -p_{0} = \alpha^{2}p^{0}+\omega L_{z}
$$
It is interesting to write these relations in terms of FIDOs 
measurement of energy and angular momentum. The generator of 
axisymmetry is unchanged: $\partial_{\phi}$; consequently the 
local measurement of the angular momentum gives the correct 
invariant $L_{z} = p_{\phi}\vert_{FIDO}$. This is different with 
energy that varies locally; and the generator of temporal invariance
$\partial_{t} = \alpha \partial_{\tau} -\omega \partial_{\phi}$. 
Therefore $-p_{0} = -\alpha p_{0}\vert_{FIDO} + \omega 
p_{\phi}\vert_{FIDO}$. Thus the energy invariant, corresponding to 
the energy measured at infinity, is given by
$$
\epsilon_{\infty} = \alpha \epsilon +\omega L_{z} \ .
$$
In the ergosphere, it is possible to get a negative energy for $L_{z}<0$.
This is at the origin of the Penrose process. Let a body in a state of positive 
energy $\epsilon_{0}$ in the ergosphere that decays into one piece with positive 
energy $\epsilon_{1}$ following an outward geodesic and another piece 
falling down with a state of negative energy $\epsilon_{2}$. Then 
$\epsilon_{1}=\epsilon_{0}-\epsilon_{2} > \epsilon_{0}$, thus the 
outgoing body emerges with more energy to the expense of the rotation 
energy of the Black Hole.

In the "3+1"-split formulation, the forces exerted on matter are 
measured by FIDOs in ${\cal F}_{3}$ (see \cite{TPD}). The gravitational interaction 
is described by a local gravity field $\vec g = -\nabla \log\alpha$ 
(times $c^{2}$ in usual units) and by a "gravito-magnetic" (GM)
field $H_{ij}$, defined by $H_{ij} = 
\frac{1}{\alpha}\nabla_{i}\beta_{j}$, which acts on a particle like a 
magnetic field: $dp_{i}/d\tau = H_{ij}p^j$. Indeed, like the magnetic 
field that complements the electric field, as required by relativity,
when charged particles are moving, similarly the GM-field complements 
the gravity field when massive particles are moving. The frame angular 
velocity $\omega$ is therefore called "gravito-magnetic" potential. 
The GM-field is responsible for a precession effect, the 
Lense-Thirring effect. Moreover, it acts on a viscous disk in such a 
way that matter is enforced to rotates in the equatorial plane 
orthogonal to the Black Hole rotation axis. Spinning black holes 
maintain a stable direction for jets that are supposed to be 
launched along the rotation axis; and if the alignement of the disk 
axis and the Black Hole axis is not completely achieved, the 
Lense-Thirring effect would produce an observable precession of the 
jets.

\subsection{The Membrane Paradigm}\label{subsec.MEMPARA}

An important issue about FIDOs is that, for BL-frame, they run faster and 
faster against the geodesic flow when closer to the horizon and all 
the motions seem to synchronise with FIDOs motions, because the 
velocity is such that $\vert \vec v \vert =\alpha^{-1}\vert 
\frac{d\vec x}{dt}+\vec \beta\vert <1$, which implies $\vert 
\frac{d\vec x}{dt}+\vec \beta\vert$ as small as $\alpha$. Deviations 
to FIDOs motions close to the horizon are very slow motions.

Transverse electric and magnetic fields diverge like $\alpha^{-1}$ 
close to the horizon and the most diverging contributions satisfy the 
property of a converging transverse electromagnetic wave in 
vacuum, namely $\vec B = -\vec n \times \vec E$ (\cite{TPD}). In the membrane 
paradigm, boundary conditions are set up at a 2-manifold defined by 
$\alpha(r,\theta) = \alpha_{H} \ll 1$. The normal electric field is 
accounted by an effective surface charge such that $\sigma_{H} = 
E_{n}/4\pi$ and a vacuum surface resistivity ($R_{H}= 4\pi = 377$ Ohms) 
accounts for a fictive current that closes the electric circuit such that 
the surface current $\vec{\cal J}_{H}$ generates the magnetic field $\vec 
B = 4\pi\vec{\cal J}_{H}\times \vec n = -\vec n \times \vec E$ with 
$\vec E = R_{H}\vec{\cal J}_{H}$.

At the membrane it is convenient to rescale all the diverging fields 
by multiplying them by $\alpha_{H}$ (\cite{TPD}). In particular, one defines the local 
surface gravity $g_{H}=\alpha_{H}g$.

An effective Ohmic dissipation is thus proposed by the membrane 
paradigm, which therefore implies a possible entropy production. 
Indeed hiding all the "sluggish" dynamics inside the membrane amounts 
to ignore a lot of informations and to constitute a state of high 
entropy. The local quantum fluctuation time scale is such that $\delta 
t \sim c/g_{H}$ and a corresponding fluctuation temperature, $T_{H} 
\sim \hbar /\delta t$, has been 
proposed by Hawking (\cite{H}): 
$$
T_{H} = \alpha_{H}T = 
\frac{\hbar}{2\pi}\frac{g_{H}}{c}=\frac{m_{P}^{2}c^{2}}{8\pi M} \ ,
$$
where $m_{P}$ is the Planck mass ($m_{P} \equiv (\hbar c/G)^{1/2}
\sim 10^{19} GeV$); this 
temperature is very low for astrophysical black holes and the 
quantum evaporation very inefficient. The entropy  
$S_{H}$ is $1/4$ times the ratio of the horizon surface area over the 
square of the Planck length (Bekenstein) \cite{BEK}: 
$$
S_{H}=2\pi(1+\sqrt{1-a_{s}^{2}})
\frac{r_{g}^{2}}{\ell_{P}^{2}}  \ ,
$$
which is an extremely large number for 
astrophysical black holes 
($\ell_{P} \equiv (\hbar G/c^{3})^{1/2} \simeq 1.6 \times 10^{-33} cm$).
Although ineffective in astrophysical applications, these concepts are 
very interesting and profound, and they are very relevant in string 
theory at Planck scale.
This entropy has recently been derived from quantum gravity theory (\cite{STRO}) by 
counting the number of quantum states compatible with 
the macroscopic parameters of a black hole.

\subsection{Electromagnetism in Kerr metric}\label{subsec.ELEKER}

In FIDO's space ${\cal F}_{3}$, electromagnetism can be described by 
local Maxwell equations with local time. Then the universal time can 
be introduced, which changes Maxwell equation by inserting the 
lapse function $\alpha$ and the GM-potential $\omega$ at the 
appropriate places. The 
electromagnetic field is thus coupled with the gravitation field. The 
coupling with matter is insured by Ohm's law (i.e. $\vec E + \vec v \times 
B = \eta \vec J$ for non-relativistic flows). In most astrophysical applications, the plasma is 
very conducting and the resistivity effect ($\eta$) is negligible 
(the Reynolds magnetic number is very large); the 
magnetic field is thus frozen in the plasma flow and the electric 
field is just there to compensate the effect of motions across the 
magnetic field, it vanishes in a frame comoving with the magnetic 
lines. 

In axially symmetric configuration, it is convenient for both 
analytical and numerical calculations to describe the 
electrodynamics with only two scalar functions $a$ and $b$, that are 
defined as follows. Any vector field $A$ can be splitted in two pieces, 
the poloidal one, $A^{r}e_{r}+A^{\theta}e_{\theta}$, and the toroidal 
one, $A^{\phi}e_{\phi}$. The basis vectors are not necessarly unitary; it 
is even convenient to use the natural basis associated with the 
coordinates and often labeled $\{\partial_{r}, \partial_{\theta}, 
\partial_{\phi}\}$ and the field components are the contravariant 
components. The covariant components are such that $A_{j} = g_{jk}A^k$. Covariant 
components change like components of a differential form under a change of coordinates, whereas contravariant components change like components of a first order differential 
operator; which justifies the notation with partial derivatives for the natural base of the vector fields; similarly the natural base of covariant field can suitably be written $\{dr, d\theta, d\phi\}$. The same when the time coordinate is introduced, one has a fourth base element $\partial_t$ for the contravariant field base and $dt$ for the covariant field base.
The poloidal magnetic field is derived from the covariant component 
of the toroidal vector potential; and one notes $a \equiv A_{\phi} = 
\bar r^{2}A^{\phi} = \bar r A_{\hat \phi}$ where $A_{\hat \phi}$ is 
the usual toroidal component for unitary basis vectors. The magnetic 
flux across an horizontal disk of constant $r$ and $\theta$ points 
is $2\pi a$.
\begin{equation}
\label{ }
\vec B_p = -\frac{\vec e_{\phi}}{\bar r} \times \nabla a
\end{equation}
\begin{equation}
\label{ }
E_{\phi}= -\frac{1}{\alpha}\frac{\partial}{\partial t}a
\end{equation}
The poloidal component of Ohm's law reads:
\begin{equation}
    \frac{\partial}{\partial t}a +\alpha v^{j}\nabla_{j}a = 0
    \label{eq:EVA}
\end{equation}
This clearly expresses the frozen in condition of the magnetic flux in 
the flow. Similarly the poloidal current is derived from the 
covariant component of the magnetic field $B_{\phi}$, noted $b$. From 
Ampere theorem in magneto-static approximation, the 
current intensity crossing the same disk as before is $I= 2\pi \alpha 
b/4\pi= \alpha b/2$. 
The toroidal part of the Faraday induction equation reads:
\begin{equation}
    \frac{\partial}{\partial t}b +\bar r^{2} div[\alpha b/\bar 
    r^{2}\vec v_{p} -(\alpha \Omega + \omega) \vec B_{p}] = 0
    \label{eq:EVB}
\end{equation}
The matter is supposed to rotate at an angular velocity $\Omega 
\equiv v^{\phi}$.

Consider a stationary magnetosphere. Relativistic MHD will be reduced to five invariant quantities
and a non-linear partial differential equation governing the transverse equilibrium of each magnetic surface, the so-called Grad-Shafranov equation (see \cite{VBP}). The magnetic surfaces are 
characterised by constant values of $a$. The toroidal electric field 
measured by FIDOs thus vanishes and the poloidal flow is along the 
poloidal field lines ($\vec v_{p} \times \vec B_{p} = \vec 0$).
The matter flux, with a bulk Lorentz factor $\Gamma$ can be written as 
follows $\alpha \rho \Gamma \vec v_{p} 
= \kappa \vec B_{p}$ and, since in stationary flow, it is divergence 
free like the magnetic field, $\vec B_{p}.\nabla \kappa = 0$, and the 
function $\kappa$, which is the flux of matter per magnetic flux, is constant along the field lines, it depends on $a$ 
only; this is the first of the five invariants:
\begin{equation}
\label{ }
\kappa(a) = \frac{d\dot M_j (<a)}{2\pi da} \ .
\end{equation}

Inserting the mass flux in term of the magnetic field in the 
induction equation (\ref{eq:EVB}), it is easily found that the 
angular velocity $\Omega_{*}$, defined by
\begin{equation}
    \Omega_{*} = \alpha \Omega + \omega -\frac{\kappa}{\rho \Gamma 
    \bar r^{2}}b \ ,
    \label{eq:OMSTAR}
\end{equation}
is also constant over each magnetic surface ($\Omega_{*}(a)$). It 
turns out that this angular velocity is the angular velocity of the 
frame in which the electric field vanishes, and one calls it the 
angular velocity of the field lines. Indeed the poloidal electric field 
is obtained from Ohm's law by inserting 
\begin{equation}
    b = \frac{\rho \Gamma \bar r^{2}}{\kappa}(\alpha \Omega + \omega 
    -\Omega_{*}) \ ,
    \label{eq:B}
\end{equation}
and one find
\begin{equation}
    \vec E = -\frac{\Omega_{*}-\omega}{\alpha c}\nabla a \ .
    \label{eq:E}
\end{equation}
A way to see that the field lines rotate despite the axisymmetry is to consider the toroidal Poynting flux which reads $\frac{(\Omega_{*}-\omega)\bar r}{\alpha}\frac{B_p^2}{4\pi}$; it clearly shows that poloidal field energy rotates with respect to ZAMOs at an angular velocity $(\Omega_{*}-\omega)/\alpha$ and $\Omega_*$ in the BL-system of coordinates.
Equation (\ref{eq:B}) expresses the current generation due to the 
shift of angular velocity across the field lines. Having got the electric 
field and the current, the main ingredients to estimate the Poynting 
flux are at hand.

\section{General Properties of Jets}\label{sec.EN}

The general properties of jets depend on the transport of the angular momentum they carry from the central rotator, spinning Black Hole, accretion disk or torus. Their power, motions and collimation are tightly related to the transport of angular momentum which is itself linked to the electric current flowing in the jet.

\subsection{Transport of angular momentum}\label{subsec.TAM}

The specific angular momentum is the toroidal covariant component of 
the specific momentum: $l\equiv u_{\phi} c = \Gamma v_{\phi} =\Gamma 
\Omega\bar r^{2}$.

Its transport equation expresses the conservation of the global 
angular momentum of the system:
\begin{equation}
    \frac{\partial}{\partial t}[(e+P)\Gamma l+ S_{\phi}^{(e.m.)}] +
    div[(e+P)\alpha \Gamma l \vec v_{p} -\frac{\alpha}{4\pi}(b\vec 
    B_{p}+E_{\phi}\vec E_{p})] = 0 \ .
    \label{eq:EAM}
\end{equation}
In stationary state ($E_{\phi}=0$), inserting the mass flux in term of 
the poloidal magnetic field leads to a third constant $l_{*}(a)$ on 
each magnetic surface that expresses the conservation of the total 
angular momentum:
\begin{equation}
    wl-\frac{\alpha b}{4\pi \kappa} = l_{*}(a) \ ,
    \label{eq:LSTAR}
\end{equation}
where $w=(e+P)/\rho c^2$ is the specific relativistic enthalpy.
It is noteworthy to remark that in the region where the magnetic field dominates (quasi force free regime)
the total specific angular momentum carried by matter is directly related to the current ($I= \alpha b/2$) :
$l_{*}(a) \simeq -I/2\pi \kappa$. For consistency the electric current must be a poloidal invariant in this regime, which is true for the force free approximation. Since $I < 0$ because generated by positive rotation with respect to the direction of the poloidal field, the specific angular momentum has the sign of mass flux ($\kappa$). Infalling matter carries negative angular momentum towards the ergosphere of the Black Hole.

\subsection{The AlfvŽn surface}

The AlfvŽn surface is one of the most important critical surface of the flow. A fast flow is supposed to pass smoothly across this surface in order to prevent backward AlfvŽn waves to accumulate there and generate a strong current perturbation. Relativistic AlfvŽn waves propagate at a velocity $V_A$ with corresponding Lorentz factor $\Gamma_A$ such that
\begin{equation}
\label{UA}
u_A^2 \equiv \Gamma_A^2 \beta_A^2 = \frac{B^2/4\pi}{e+P} \ .
\end{equation}
A local AlfvŽn Mach number $m$ can be defined such as $m^2 \equiv \frac{\Gamma^2 v_p^2}{c^2 u_{Ap}^2}$, where $u_{Ap}$ involves the poloidal magnetic field only. The AlfvŽn surface is such that $m^2(r, \theta) = 1$. This criticality appears when the current and the angular momentum are derived from both induction equation (\ref{eq:B}) and angular momentum equation (\ref{eq:LSTAR}). Indeed the two following relations are deduced :
\begin{equation}
\label{BM}
\frac{\alpha}{4\pi \kappa}(m^2 -1)b = l_*-\frac{w}{\alpha}(\Omega_*-\omega)\Gamma \bar r^2 \ ,
\end{equation}
\begin{equation}
\label{LM}
w(m^2-1)l = m^2 l_* -\frac{w}{\alpha}(\Omega_*-\omega)\Gamma \bar r^2
\end{equation}
The regularity conditions imply that the magnetic surfaces must cross the AlfvŽn surface at a distance from the axis such that
\begin{equation}
\label{RA}
\bar r^2 = \bar r_A^2 \equiv \frac{l_*}{(\Omega_*-\omega)\Gamma w/\alpha \vert_A}
\end{equation}
The values of $b$ and $l$ at the AlfvŽn surface are then obtained as the ratio of the derivatives along the poloidal field of the vanishing factors (l'H\^opital's rule).

The important aspect of the AlfvŽn surface is that it marks a transition between a regime of acceleration in widening magnetic surfaces and an asymptotic regime of quasi-ballistic motions, and also a transition in the transfer of angular momentum. Indeed, at the beginning of the flow where $m^2 \ll 1$ (sub-AlfvŽnic regime) and $\bar r \ll \bar r_A$, equation (\ref{LM}) shows that $l \ll l_*$ and the contribution of the magnetic torque to the angular momentum invariant $l_*$ dominates. In the opposite super-AlfvŽnic case ($m^2 \gg 1$), with $\bar r \gg \bar r_A$, $l \rightarrow l_*$, although the angular velocity decreases. In fact $l$ is already close to $l_*$ at the AlfvŽn surface.

A determination of the AlfvŽn radius is crucial but it depends on technical details. A natural estimate can be found in non-relativistic regime, where it essentially depends on the mass flux $\dot M_j(a)$ inside a flux tube defined by $a$, on the field line rotation $\Omega_*$ and the magnetic flux $2\pi a$. Assume $\Omega_* \bar r_A$ of the order of the AlfvŽn velocity at the AlfvŽn surface, then, since the kinetic and magnetic energy density are equal at this surface, the  energy fluxes are the same, and one can write:
\begin{equation}
\label{ }
{1\over 2} \dot M_j (a) \Omega_*^2r_A^2 = \lambda_0 {B_p^2 \over 4\pi} \Omega_*r_A \pi r_A^2 =
\lambda \frac{\pi^2 a^2}{r_A}  \Omega_* \ ,
\end{equation}
where $\lambda_0 \sim \lambda \sim 1$. Therefore the following estimate is derived
\begin{equation}
\label{ }
r_A = (\frac{2\lambda \pi^2 a^2}{\dot M_j \Omega_*})^{1/3} \ .
\end{equation}

A relativistic estimate will be derived in a subsection (\ref{subsec.AEF}).

\subsection{Energy transport, Bernoulli invariant}\label{subsec.BI}

In stationary state, the conservation of energy is expressed by a 
divergence free flux:
\begin{equation}
    div(\alpha^{2}\vec S) + \frac{1}{2}\alpha^{2}(\nabla_{j}\beta_{k}+
    \nabla_{k}\beta_{j})T^{jk} = 0
    \label{eq:ENF}
\end{equation}
The multiplication of the local energy flux by $\alpha^{2}$ is 
explained as follows: one factor comes from the time rescaling, another
one comes from the conversion of the energy locally measured by FIDOs 
to the energy-at-infinity. The poloidal contribution of matter is
\begin{equation}
    \vec S^{(m)}=(e+P)\Gamma^{2}\vec v_{p} \ ;
    \label{eq:SM}
\end{equation}
the contribution of the electromagnetic field is 
\begin{equation}
    \vec S^{(e.m.)}=\frac{\vec E\times \vec 
    B_{\phi}}{4\pi}=-\frac{\Omega_{*}-\omega}{\alpha}\frac{b}{4\pi}
    \vec B_{p} \ ;
    \label{eq:SEM}
\end{equation}
The second term of (\ref{eq:ENF}) is a non-trivial contribution of the 
gravito-magnetic field. It can be cast in the form of a divergence of 
the following field:
\begin{equation}
    \alpha^{2}\vec S^{(g.m.)} = \kappa l_{*}\omega \vec B_{p} = 
    (w\kappa l-\frac{\alpha b}{4\pi}) 
    \omega \vec B_{p} \ .
    \label{eq:SGM}
\end{equation}
It is important to realise that, since $b<0$ and $\Omega_{*}<\omega$ 
close to the Black Hole, the Poynting Flux is converging. But the 
contribution of the gravito-magnetic flux cancels the 
$\omega$-terms, so that the sum of electromagnetic and 
gravito-magnetic fluxes makes an outgoing energy flux. Which means 
that the gravito-magnetic effect is the one which makes an energy 
outflux from the rotating black hole. In particular, as long as the
matter contribution is negligible (like in force free regime), the 
energy flux is
\begin{equation}
    \alpha^{2}\vec S^{(e.g.m.)} = -\frac{\alpha b}{4\pi}\Omega_{*}
    \vec B_{p} = \frac{-I}{2\pi}\Omega_* \vec B_p \ .
    \label{eq:SEGM}
\end{equation}
However as one looks at a remote distance from the black hole, the GM-potential 
vanishes and the outgoing energy flux becomes essentially 
the Poynting flux. The form of the GM-energy flux, involving the 
flux of angular momentum and the space drag effect, is reminiscent of the 
Penrose effect. Indeed, the energy flux carried by the GM-field is 
nothing but the flux of angular momentum times $\omega$, which 
completes the usual energy flux measured locally, as seen previously in subsection (\ref{subsec.SPLIT}). 
Infalling matter ($\kappa<0$) carries a negative specific angular momentum
($l_{*}<0$), which, 
with the frame dragging, produces an outward energy flux.

Following Michel \cite{M1}, it is convenient to define the "magnetisation parameter" $\sigma$ as the ratio of the Poynting flux over matter energy flux:
\begin{equation}
\label{SIG}
\sigma \equiv \frac{(\vert I \vert /2\pi)\Omega_* d\Phi}{\alpha w \Gamma d\dot M_jc^2}  = \frac{\vert b \vert \Omega_*/4\pi}{w \Gamma \kappa c^2} \ .
\end{equation}
This number must be large at the beginning of the jet in order that the Poynting flux propells matter efficiently and then it diminishes in favour of the matter energy flux. The magnetization takes on a simple form at the AlfvŽn surface:
\begin{equation}
\label{SA}
\sigma_A = \frac{\Omega_*(\Omega_*-\alpha \Omega -\omega)\bar r_A^2}{\alpha^2 c^2}\mid_A \ ;
\end{equation}
Far enough from the Black Hole, one gets $\sigma_A \simeq \Omega_*^2 \bar r_A^2/c^2$, and an AlfvŽn surface far enough from the light cylinder would be expected, otherwise, the jet acceleration would almost end there. 

From the divergence free condition when radiation and dissipation can be neglected, a Bernoulli invariant can be derived (the fourth of the family):
\begin{equation}
    \alpha w \Gamma c^2 -\frac{\alpha b}{4\pi \kappa} \Omega_{*} +w\omega 
    l \equiv {\cal B}(a) \ .
    \label{eq:BINV}
\end{equation}
Starting with $\Gamma_0 = (1-\Omega_0^2 \bar r_0^2/c^2)^{-1/2}$, ${\cal B}(a) = \alpha_0 w_0 \Gamma_0(1+\sigma_0) + w_0 \omega l_0$, with an initial magnetisation $\sigma_0$ which is expected to be large in order that the magnetic field drive a relativistic flow. It is often convenient to choose these "initial" values at the light cylinder beyond the "watershed" surface.
From (\ref{eq:BINV}), it follows that, at remote distance from the Black Hole, the product $\Gamma (1+\sigma)$ is almost constant on each magnetic surface; and since $\Gamma \sigma = k\vert I \vert$ (where $k$ is a constant), the Bernoulli invariant implies that $\Gamma + k \vert I \vert$ is almost constant on each magnetic surface. The expected growth of the bulk Lorentz factor requires a strong decrease of the electric current intensity, which cannot be described in force free MHD where the current intensity is constant on each magnetic surface.

The decrease of the magnetization by a decrease of the current and an increase of the motions can be translated in angular momentum language, because of the ideal MHD assumption. Indeed the evolution of the magnetization parameter is described in term of the evolution of the angular momentum, by inserting the angular momentum equation (\ref{eq:LSTAR}):
\begin{equation}
\label{ }
\sigma = \frac{\Omega_*(l_*-wl)}{\alpha^2 \Gamma c^2} \ ,
\end{equation}
the large initial value corresponds to a high value of the magnetic breaking effect $\Omega_* l_*$, with $l \ll l_*$, and the decrease at large distance where $w \simeq 1$ corresponds to an increase of the matter angular momentum in widening magnetic surfaces that tends to $l_*$.

Far from the Black Hole, the Bernoulli function takes on a simple form, even after inserting the expressions for $b$, eq: (\ref{BM}), and $l$, eq:(\ref{LM}):
\begin{equation}
    {\cal B}(a) = w\Gamma c^2 - \frac{l_*\Omega_* - w\Gamma \Omega_*^2 \bar r^2}{m^2-1} \ .
    \label{eq:BINVF}
\end{equation}
the bulk Lorentz factor is then derived for  prescribed magnetic surfaces :
\begin{equation}
\label{GAM }
(m^2-1 + \xi^2)w \Gamma = (\sigma_0 + w_0)m^2 - w_0 \ ,
\end{equation}
where $\xi \equiv \Omega_* \bar r/c$. We will discuss the asymptotic evolution of the Lorentz factor in the next subsection.

The location of the AlfvŽn surface is constrained by setting $m^2 = 1$, which leads to
\begin{equation}
\label{LOCA}
\frac{\Omega_*^2 \bar r_A^2}{c^2} w_A \Gamma_A = \sigma_0 \ .
\end{equation}
For a not very hot plasma $\bar r_A \simeq \bar r_L (\sigma_0 /\Gamma_A)^{1/2}$ ; at this stage however the value of $\Gamma_A$ is unknown.
Most of the Poynting flux would have been converted into matter energy flux at the AlfvŽn surface and thus acceleration of the flow almost end there... Therefore the flow would be relativistic only if the AlfvŽn velocity is relativistic, which amounts to the condition of a very dominating magnetic pressure there: $B_p^2/8\pi \gg e+P$.

The non-relativistic limit ($\alpha \simeq 1+ \Phi/c^2$ and $w \simeq 1 + h/c^2$) of this invariant leads to
\begin{equation}
    \frac{1}{2}\vec u^{2}+\Phi + h -\frac{b\Omega_{*}}{4\pi 
    \kappa} = {\cal B}_{nr}(a) \ ,    
\label{eq:BINVC}
\end{equation}
where the mass contribution has been withdrawn. This is not necessary from a formal point of view, but convenient for the understanding of the difference between the relativistic and the non-relativistic regimes. Indeed this governs the motions along the poloidal magnetic field lines and particularly the increase of kinetic energy, which is easier than the increase of mass-energy in the relativistic regime  for the Poynting flux. It is noteworthy to remark that a magnetisation parameter can also be defined in the non-relativistic MHD where energy mass is disregarded. That non-relativistic magnetisation can be made more easily large than in the relativistic case, which makes non-relativistic flow easier to be accelerated, moreover they do not suffer any radiation drag... But similarly to the relativistic regime, the acceleration of the flow is almost ended at the AlfvŽn surface where $\sigma^{nr}_A = 2\Omega_*^2 \bar r_A^2/V_A^2$. Indeed $V_A/\Omega_*$ is the typical scale of the MHD flow and the AlfvŽn radius takes on values of this order.

A fifth invariant is the plasma entropy ($P \rho^{-\gamma}= K(a)$).
To close the system of equations that entirely describes the jet 
formation, an equation that determines the magnetic surfaces by 
stating the transverse balance is needed. This is the Grad-Shafranov 
equation (\cite{FN} \cite{OI}). These six equations determine $a$, $b$, $l$, $u$, $\rho$, $P$; the five invariants must be specified at some surface crossing all the relevant magnetic surfaces, e.g. at the AlfvŽn surface or  the surface of the rotor, and two boundary conditions must be set for the integration of Grad-Shafranov equation.
Once a jet is launched, it has to cross three critical surfaces (see \cite{S} and \cite{VTT} for a good analysis of the critical surfaces). In 
the case of the pulsar wind, the matter is so tiny compared to the 
intensity of the magnetic field, that only the light cylinder at 
$\bar r_{L} \equiv \Omega_{*}/c$ requires a regularity condition. In 
AGNs, micro-Quasars and GRBs, matter pressure is not negligible compared to magnetic pressure. 
The Alfv\'en surface replaces the light cylinder and 
there are, in addition, slow and fast magnetosonic surfaces.

\subsection{Asymptotic evolution of the flow}\label{subsec.AEF}

From Bernoulli equation, the asymptotic evolution of the Lorentz factor of the flow can easily be derived; it is governed by the following equation when the flow has a non-relativistic temperature:
\begin{equation}
\label{EVG}
\Gamma^3 -(1+\sigma_0-\sigma_c + \frac{\sigma_c}{\xi^2}) \Gamma^2 + \frac{\sigma_0}{2} = 0 \ .
\end{equation}
In eq.(\ref{EVG}), the parameter $\sigma_c$ corresponds to a kind of modified value of the magnetisation parameter that depends on the local poloidal field and is defined by
\begin{equation}
\label{SIGC }
\sigma_c \equiv \frac{\Omega_*^2 B_p r^2 \sin^2 \theta}{4\pi w \kappa c} \ .
\end{equation}
This parameter is close to the value of the magnetisation when the asymptotic magnetic configuration is described by a monopolar field:
\begin{equation}
\label{SMO}
\sigma_{mono} = \frac{\Omega_*^2 B_r r^2 \sin^2 \theta}{4\pi w \kappa c} \ ,
\end{equation}
in which $B_r r^2 = \, contant$, thus leading to a constant magnetisation on the magnetic surfaces (cones in this approximation) like in \cite{M1}. Clearly, a monopole approximation that would have been assumed too early in the description of the flow (this is an exact solution of the complete set of equations (\cite{M2}, \cite{BR}) would prevent any significant acceleration since $\sigma_c \simeq \sigma_0$. Moreover an asymptotic monopole solution cannot exist if $\sigma_c = \sigma_0$ as can easily be seen on eq.(~\ref{EVG}), that would have no root for $\bar r \rightarrow \infty$. Thus the difference $\Delta \sigma \equiv \sigma_0-\sigma_c$ must be carefully treated in the flow description.

This important equation eq.(~\ref{EVG}) has been obtained by inserting $m^2 \equiv u^2/u_A^2 = 
(u/\sigma_c ) \xi^2$ and $\Gamma^2 = 1+u^2 + l^2/c^2\bar r^2$. For large $\Gamma$, it turns out that the approximation $u \simeq \Gamma - 1/2\Gamma$ is sufficient.

The fast magneto-sonic surface is another important critical surface located beyond the AlfvŽn surface, but usually considered in the literature as not much farer. It can be found from eq.(~\ref{EVG}) which indicates that the derivative of $\Gamma$ along the magnetic line has a denominator that vanishes when $\Gamma = \Gamma_F$ with $\xi = \xi_F$ such that
\begin{equation}
\label{GF}
3\Gamma_F^2 = 2(1+\Delta \sigma_F + \sigma_c/\xi_F^2) \ .
\end{equation}
The Lorentz factor at this surface must always takes the value $\Gamma_F = \sigma_0^{1/3}$ (after having inserted eq.(\ref{GF}) into eq.(\ref{EVG}) and the radius is such that
\begin{equation}
\label{RF}
\bar r_F^2 = \bar r_L^2 \frac{\sigma_c}{\frac{3}{2} \sigma_0^{1/3} -1-\Delta \sigma} \ ,
\end{equation}
which shows that the FM-surface exists only if $\Delta \sigma < \frac{3}{2} \sigma_0^{1/3} -1$. Therefore $\bar r_F \sim \bar r_L \sigma_0^{1/3}$. Since $\Gamma_A < \Gamma_F$, it turns out that $\bar r_A 
\sim \bar r_L \sigma_0^{1/3}$ also; these two estimates are general results for a cold flow.

At this stage, a preliminary discussion can be addressed. 
\begin{itemize}
  \item If $\Delta \sigma \sim \sigma_0^{1/3}$ even far beyond the FM-surface, then $\Gamma_{\infty} \sim \sigma_0^{1/3}$ and most of the Poynting flux has not been converted into jet power. The asymptotic solution can be of monopolar type.
  \item If $\Delta \sigma_{\infty} \sim \sigma_0$ (which would necessarily held far beyond the AlfvŽn surface only), a much better conversion would occur and $\Gamma_{\infty} \sim \sigma_0$. Estimating $\sigma_0$ at the light cylinder, the ratio between both magnetisations is
\begin{equation}
\label{SOC}
\frac{\sigma_0}{\sigma_c} = \frac{\vert B_{\phi}\vert_L}{\Gamma_LB_p}\frac{\bar r_L^2}{\bar r^2} = \vert \frac{B_{\phi}}{\Gamma B_p}\vert_L \frac{B_p \bar r^2\mid_L}{B_p \bar r^2} = \frac{\Omega_*-\Omega_L}{\Omega_*u_L} \frac{B_p \bar r^2\mid_L}{B_p \bar r^2}  \ .
\end{equation}
This ratio clearly tells how an efficient conversion of the Poynting flux into jet power could occur. The high intensity of the toroidal field compared to the poloidal one during the non-relativistic stage of acceleration is favourable, especially if the poloidal velocity $u$ is still subrelativistic at the light cylinder ($u_L <1$). But this is spoiled by the condition for smooth crossing the FM-surface (however see next paragraph where the case of a warm plasma is discussed). The other favourable circumstance  is based on the possibility that $ B_p^2 r^2 \sin^2 \theta$ becomes sufficiently smaller than its value at the light cylinder. This is the attempt followed by N. Vlahakis (\cite{VL}).
  \item If the relativistic flow is collimated by a non-relativistic jet, such that $\bar r_{\infty} > \bar r_L \sigma_0^{1/3}$, and $r^2B_p^2\sin^2 \theta < r^2B_p^2\mid_L \sin^2 \theta_L$, then $\Delta \sigma_{\infty} \sim \sigma_0$. Therefore the flow undergoes a sizeable conversion  of the Poynting flux into jet power. 
\end{itemize}

{\it In the case of a warm flow}.\\

Crossing the FM-surface seems quite constraining for a flow having a non-relativistic temperature. From eq:(\ref{GAM}), it is not difficult to modify the evolution equation of the bulk Lorentz factor for a flow that has a relativistic temperature, so that $w = 4 \bar \gamma/3 > 1$. It becomes
\begin{equation}
\label{ }
w\Gamma^3 -(w_0 +\sigma_0-\sigma_c w (1-{1\over \xi^2})) \Gamma^2 + \frac{\sigma_0+w_0}{2} = 0 \ .
\end{equation}

One obtains the modified value of the Lorentz factor at the FM-surface $\Gamma_F= (\frac{\sigma_0 + w_0}{w_F})^{1/3}$ and the location of the FM-surface is such that
\begin{equation}
\label{ }
\bar r_F^2 = \bar r_L^2 \frac{w_F \sigma_c}{{3\over 2} w_F^{2/3}(\sigma_0+w_0)^{1/3}-w_0+w_F \sigma_c-\sigma_0}
\end{equation}
A relativistic temperature at the FM-surface greatly helps its smooth crossing,  even for $w_0 \ll \sigma_0$, provided that $w_F >1$ so that $w_F\sigma_c > \sigma_0$. Then the FM-surface is located at a few light cylinder radii with 
\begin{equation}
\label{ }
\bar r_F^2 = \bar r_L^2 \frac{w_F \sigma_c}{w_F \sigma_c-\sigma_0}
\end{equation}

But the flow must have cooled at remote distance such that  $w_{\infty} \simeq1$ and $\Delta \sigma_{\infty} \sim \sigma_0$ to allow a large conversion such that $\Gamma_{\infty} \sim \sigma_0$. Thus the recipe to get a successful conversion of power is an initial magnetisation $\sigma_0$ significantly larger than its monopole estimate, thanks to the non-relativistic stage of acceleration, together with a relativistic warm plasma up to the FM-surface. But the hotter the plasma the weaker the magnetisation; indeed for a relativistic plasma in mildly relativistic motions at the light cylinder, $\sigma_0 \sim B_p^2/8\pi P$. In particular at equipartition the magnetisation parameter is of order unity and thus the formation of a relativistic jet becomes difficult. However a thermal relativistic wind is set up; this is the assumption of the "fire ball" model of GRBs (\cite{RM}) despite the very high intensity of the equipartition magnetic field that can reach $10^{15}G$... Nevertheless there also exist some theoretical attempts to explain GRB relativistic wind with the electromagnetic formalism presented in this section  (as for instance \cite{VK}).

It is quite conceivable that a cold flow with a high initial magnetisation can be entrained by a large Poynting flux without succeeding to establish a stationary flow that crosses the critical surfaces. This would give rise to Poynting flux dominated flow with a very large Lorentz factor for the electromagnetic field $\Gamma_{e.m.} \gg 1$ with some amount of perturbing matter having $\Gamma < \Gamma_{e.m.}$ (Lyutikov \& Blandford \cite{LB}) . 

\subsection{Blandford-Znajek mechanism}\label{subsec.BZ}

The celebrated Blandford-Znajek mechanism (\cite{BZ}) consists in taping the 
rotation energy of a Kerr Black Hole. From thermodynamical arguments, 
the energy-mass of the black hole is shared in an irreducible part 
related to the entropy and a rotational part that can be extracted. 
At maximum rotation ($a_{s}=1$), $M_{irr}=\frac{\sqrt{2}}{2}M$ and 
$M_{rot}=(1-\frac{\sqrt{2}}{2})M \simeq 0.29\, M$. As seen previously, 
the energy extraction is done by the gravito-magnetic field ("space 
drag"), through a kind of Penrose effect. In the neighbourhood of the 
horizon, the effect of the magnetic field is supposed to dominate the 
inertial effect and a force free approximation can be assumed. The 
poloidal current is thus aligned with the poloidal magnetic field and 
therefore the current intensity is conserved on each magnetic 
surface: $I(a)$. Since the electric field is known, close to the horizon, 
the current is determined by the converging wave condition in the 
vacuum, as stated in subsection (3.2), namely $\vec B^{trans} = 
-\vec n \times \vec E(1+ {\cal O}(\alpha))$, or 
$B_{\hat \phi} =-E_{\hat \theta}$ and 
$B_{\theta} = 0$. From (\ref{eq:E}), 
one straightforwardly gets
\begin{equation}
    b =-\frac{\Omega_{H}-\Omega_{*}}{\alpha c}\bar r^{2}B_{n}
    \label{eq:BHOR}
\end{equation}
In agreement with the statements of subsection (3.2), 
both the transverse electric field and the toroidal magnetic field 
diverge like $\alpha^{-1}$ and contribute to the energy flux at 
${\cal O}(1)$. This form of $b$ compared to (\ref{eq:B}) indicates that 
there is necessarly a constant flux of falling matter on the polar 
cap of the Black Hole such that $\alpha \rho \Gamma = -\kappa^{-} B_{n}$ 
($\kappa \mapsto -\kappa^{-} < 0$) and $v_p = -1 + {\cal O}(\alpha)$. 
The total outgoing energy flux in the vicinity of the horizon is thus:
\begin{equation}
    P_{BZ} \equiv \int_{\cal H} \alpha^{2} \vec S.\vec n d^{2}\sigma =
     \int_{\cal H} \Omega_{*}(\Omega_{H}-\Omega_{*}) \bar r^{2}
     \frac{B_{n}^{2}}{4\pi c} d^{2}\sigma \ .
    \label{eq:PBZ}
\end{equation}
This corresponds to the loss of Black Hole energy-mass per unit time in the form of rotation energy.
This power is maximum when $\Omega_{*} = \Omega_{H}/2$, which can be 
interpreted as an impedance optimisation (\cite{TPD}). The electric circuit is 
supposed to be loaded far from the black hole by a dissipative region 
with a resistance equal to the resistance of the black hole, i.e. 
vacuum resistance. Some criticism of this optimisation has been addressed by 
Punsly and Coroniti (\cite{PU} \cite{PC}), who argued that there is no causality relation between the flow beyond the matter source around the watershed surface and the flow falling toward the horizon and the ergosphere. However numerical simulations (\cite{CMc} \cite{CMd}) have shown the capability of the process to generate relativistic jets, as long as no coupling with a radiation field is taken into account.

As shown previously, the Poynting flux converges 
towards the Black Hole: as usual the Poynting flux converges towards 
the ohmic dissipation region, where the electric field and the current 
point towards the same direction. The dissipation of the electromagnetic 
field in the Black Hole is given by:
\begin{equation}
    P_{dis} \equiv \int_{\cal H}(\Omega_{H}-\Omega_{*})^{2}\bar r^{2} 
     \frac{B_{n}^{2}}{4\pi c}d^{2}\sigma \ .
    \label{eq:PDIS}
\end{equation}
This corresponds to the increase of Black Hole entropy as previously introduced (in subsection (3.2)), i.e.
$P_{dis} = T_H dS_H/dt$. At the "impedance optimisation", $P_{BZ} = P_{dis}$.
At a few gravitational radii, the space dragging decreases enough so 
that $\omega < \Omega_{*}$, the electric field changes its direction, 
and the Poynting flux dominates, carrying energy-momentum far away 
and eventually gives it up to matter. This is potentially able to 
generate a relativistic jet of power:
\begin{equation}
    P_{BZ} \simeq 10^{42} a_{s}^{2}(\frac{B}{10^{4}G})^{2}
    (\frac{M}{10^{8}M_{\odot}})^{2} \, ergs/s \ .
    \label{eq:PBZC}
\end{equation}
It is interesting to compare it to Eddington luminosity and also to bear in 
mind the Eddington scaling of the magnetic field:
\begin{equation}
    \frac{P_{BZ}}{L_{Ed}} \simeq 10^{-4} a_{s}^{2}(\frac{B}{10^{4}G})^{2}
    (\frac{M}{10^{8}M_{\odot}}) \simeq 10^{-4} a_{s}^{2}
    (\frac{B}{B_{*}})^{2} \ .
    \label{eq:PSL}
\end{equation}
This ratio is independent of the Black Hole mass and it is not expected 
that the magnetic field could be larger than ten times $B_*$, the Eddington estimate for $a_s = 0$. 
It clearly shows that the Blandford-Znajek mechanism has a relatively significant power only for 
a central engine that works well below the Eddington luminosity, as would be the case of a 
weakly radiating torus.

\section{Relativistic and non-relativistic jets}\label{sec.RNRM}

How a Poynting flux is  converted into energy-momentum of matter? 
This is a difficult problem that has not been solved yet, although there are some possibilities previously discussed. This flux 
propagates through a plasma which, first, is falling ($\kappa <0$) on the 
Black Hole, then is generated in the form of a pair plasma around a stagnation 
surface, and finally expelled ($\kappa >0$) (see figure (1)). Looking at 
the Bernoulli invariant (\ref{eq:BINV}), it can be seen that the 
asymptotic bulk Lorentz factor is directly determined by the value of 
the Bernoulli function, especially if the current decreases along the 
jet, which is likely. As previously seen, to get a high Lorentz factor, it is necessary to have a large 
magnetisation, which means a strong contribution of the magnetic field to the angular momentum transfer, at the beginning of the jet since, 
\begin{equation}
\label{BER}
{\cal B}(a) = \alpha w \Gamma c^2 (1+\sigma) + w \omega l
\end{equation}
Moreover the strength of the magnetisation is nothing but the strength of the current 
(or toroidal field) at the beginning of the jet, which constrains the angular momentum 
invariant. It is thus expected that there exists a starting region of the jet fairly free from gravitation, such 
that the total angular momentum is dominated by the field, i.e. $\vert b \vert/4\pi \kappa \simeq l_*$, and $\sigma \simeq \sigma_* \equiv l_*\Omega_*/c^2$.
Thus the conversion of the Poynting flux would be successful and $c^{2}\Gamma_{\infty} \simeq {\cal B}(a)$ with ${\cal B} \simeq l_{*}\Omega_{*}$. This 
is actually its maximum possible value; however no complete analytical nor numerical solutions have been derived so far displaying such achievement. This is probably due to the defect of the "monopole" type solution that has been used as a starting solution, because it is an exact solution of the complete set of equations (as explained in subsection (4.4) and also N. Vlahakis's chapter). 

Similarly, in the 
non-relativistic case, the asymptotic bulk velocity is approximately given 
by the value of the Bernoulli function: $u_{\infty} \simeq \sqrt{2{\cal B}_{nr}}$. Similarly to the relativistic case, the Bernoulli function is determined close to the source 
where the magnetic field is expected to dominate: ${\cal B} \simeq -\Omega_{*}\frac{\alpha 
b}{4\pi \kappa}\vert_{0} \simeq \Omega_*l_*$. Inserting the angular momentum equation (\ref{eq:LSTAR}) in the non-relativistic Bernouilli invariant (\ref{eq:BINVC}), one gets :
\begin{equation}
    \frac{1}{2}\vec u^{2}+\Phi + h +\Omega_*(l_*-l) = {\cal B}_{nr}(a) \ ,    
\label{eq:BINVL}
\end{equation}
Since $l \rightarrow l_*$, the same 
result is obtained for both relativistic and non-relativistic cases: ${\cal B}_{nr} \simeq l_{*}\Omega_{*}$.
In both case, the matter angular momentum starts with a low value compared to the toroidal field contribution and then increases whereas the toroidal field contribution decreases.
Therefore the relativistic flow is achieved if $l_{*}\Omega_{*} > 
c^{2}$ and $\Gamma_{\infty} \sim l_{*}\Omega_{*}/c^{2}$; if not, a 
non-relativistic flow is obtained with 
$u_{\infty} \simeq \sqrt{2l_{*}\Omega_{*}}$.
This seems simple, but the way these invariants $l_{*}$ and 
$\Omega_{*}$ are determined is not simple.

\subsection{Compton Drag}

The main danger for the relativistic flow is the Compton drag. It is 
easy to estimate the characteristic length of relaxation of  a cold 
relativistic flow that suffer energy-momentum loss by Compton 
radiation. Indeed, assume a mass flux $\dot M_{j}$ of the relativistic 
flow, composed of electrons, positrons and protons satisfying electric 
neutrality, with an electron density $n_{e} \equiv n$ (comoving) , a positron density $n^{+}= (1-\xi)n$ and 
a proton density $n_{p} =\xi n$. The mass flux across $dA$ is 
$$
d\dot M_{j}= ((2-\xi)m_{e}+\xi m_{p}) n\Gamma \beta c dA
$$
The energy decrease due to Compton drag can be described by the following equation:
\begin{equation}
\label{ }
div(\vec S^{m} + \vec S^{e.m.g.}) = -\frac{4}{3}(2-\xi)n\sigma_T c U_{\gamma} \Gamma^2<\gamma^2> \ ,
\end{equation}
where $\gamma$ is the Lorentz factor of an electron measured in the co-moving frame
and $U_{\gamma}$ is energy density of the radiation field. This expression of the Compton drag is obtained with the approximation of an isotropic radiation field; anisotropy effects will be analysed in a specific subsection (7.1) devoted to the Compton rocket effect, which necessarily requires a kinetic description.
A relaxation length $\ell_{rel} (r)$ can be properly defined such that
\begin{equation}
\label{LR}
div [(1+\sigma) \vec S^{m}] = - (1+\sigma) S^{m}/\ell_{rel} \ .
\end{equation}
One finds
\begin{equation}
\label{ }
\ell_{rel} (r) = \frac{1+\sigma}{\ell_c} \frac{e+P}{(2-\xi)nm_e c^2<\gamma^2>} r \ ,
\end{equation}
where the compactness $\ell_c$, as introduced in first section, is the main parameter that must be compared with the magnetisation $\sigma$.
The relativistic jet is killed by the Compton drag if $\ell_{rel} < r$. Two important cases must be considered: the electron-positron jet and the electron-proton jet.

In the case of a pair jet, 
\begin{equation}
\label{LRP}
\ell_{rel} = \frac{1+\sigma}{\ell_c} \frac {<\gamma>}{<\gamma^2>} r \ .
\end{equation}
as can be seen, a hot distribution of pairs makes the Compton drag more efficient than a cold distribution. The relativistic jet can be launched only if $\sigma_0 \gg \ell_c(r_0) <\gamma^2>/<\gamma>$. This constraint is severe and usually not achieved. Indeed, as previously seen for a luminosity smaller but still of the order of Eddington luminosity, the compactness $\ell_c = 10^2-10^3$. It turns out that pair creation can be catastrophic with high compactness and unavoidable in situ reacceleration of them. Therefore they get a pressure that easily reaches the magnetic pressure, which makes $\sigma \sim 1$. The relativistic jet  immediately relaxes by emitting hard X-rays.

Note that there is an amazing regime when $\sigma_0 \gg 1$ and $\sigma_0^{1/3} < \ell_c < \sigma_0$, that has been pointed out by Begelman et al. (\cite{LBC}) and Beskin et al. (\cite{VS}), because of a non-trivial effect of the magnetic field line curvature. In this regime, although the Compton drag reduces the power of the jet, the bulk Lorentz factor is increased compared to the Compton drag free case.

It is often thought that a large proton load of the flow could help, but even if we assume that there are as many protons than electrons at the origin of the jet, the relaxation length is such that:
\begin{equation}
\label{LRM}
\ell_{rel} \simeq \frac{1+\sigma}{\ell_c} \frac {m_p<\gamma_p>}{m_e<\gamma^2>} r \ .
\end{equation}
The mass ratio makes it more favourable, but still a hot distribution of electrons makes launching difficult.
Moreover protons need to be accelerated before competing with the Compton drag. In the next subsection,  the proton entrainment will be estimated. Anyway the important conclusion that can be drawn about the possibility of jet production by the Blandford-Znajek mechanism is that it requires a low compactness, which is possible only if the accretion flow is not radiatively efficient ($L < 10^{-3} L_{Edd}$).

\subsection{Baryonic pollution of a pair jet}

Diffusion of protons from the accretion disk or torus or from the non-relativistic outflow towards the pair jet takes a long time, even if we consider the fastest diffusion coefficient across the magnetic surfaces, namely the Bohm coefficient ($D_{\perp} = {1\over 3} vr_L$). Remark that the thermal estimate of Bohm diffusion is the same for both electrons and protons if they have the same temperature ($cT/eB$). Assume a constant flux of baryonic matter crossing the magnetic layer of radius $a$ and length $\Delta r$ that interacts with the leptonic flow that fastly entrains it. Then in permanent regime, the ratio of the number density of polluting protons over the external protons is given by
\begin{equation}
\label{NPOL}
\frac{n^{int}}{n^{ext}} = \frac{2D_{\perp} \Delta r}{\Gamma v_p a^2} = \frac{t_{transit}}{t_{diffus}} \ ,
\end{equation}
where $v_p = \beta_p c$ is the poloidal velocity of the jet.
This can be rewritten in term of the travel length $\Delta r_{mix}$ required to get full diffusion, $n^{int} = n^{ext}$:
\begin{equation}
\label{RMIX}
\Delta r_{mix} = \frac{3}{2}  \frac{\Gamma c}{\beta_p V_A}  \frac{a}{r_0}  a \ ,
\end{equation}
where $r_0 \equiv V_A/\omega_{cp}$; that kind of typical Larmor radius is the minimum scale of collisionless MHD description. It turns out that $a/r_0$ is a huge number (which, by the way, validates collisionless MHD in these topics), indeed
\begin{equation}
\label{LMHD}
\frac{r_g}{r_0} \simeq 0.6 \times 10^{12} \frac{M}{10^8 M_{\odot}} (\frac{n}{10^{12} cm^{-3}})^{1/2} \ .
\end{equation}
The density ratio is thus
\begin{equation}
\label{BPOL}
\frac{n^{int}}{n^{ext}} = \frac{\Delta r}{\Delta r_{mix}}
\end{equation}
For instance at $10^3 r_g$, this ratio is less than $10^{-10}$ (the ambient medium has a density much smaller than $10^{12} cm^{-3}$, $10^9 cm^{-3}$ say), the entrained proton density is $\sim 10^{-1} cm^{-3}$, whereas the pair density is  $\sim 4 \times 10^3 cm^{-3}$ (when the pair pressure is in equipartition with a magnetic field of $1G$ and an average energy  $\bar \gamma m_e c^2$ with $\bar \gamma = 10^3$). 

This issue of baryonic pollution of a relativistic pair flow is particularly important in the physics of GRBs also. In the fire ball model, the asymptotic bulk Lorentz factor is determined by the amount of baryons that have been entrained by the $e^+-e^-$- wind; and the model suffers from requiring a fine tuned value of the baryon parameter to account for the observations, without any physical argument.

Note that the pair flow is first polluted by the most energetic protons that have a larger diffusion coefficient through their larger Larmor radii.

Another way to accelerate a relativistic pair jet will be presented in a next section devoted to the "two-flow" model, where the Compton rocket effect in a hot pair plasma is taken into account.

\subsection{Confinement}\label{subsec.CONF}

Because of the toroidal field, it may be expected that the flow tends to self-collimate, if the 
magnetic tension compensates the outward pressure force and the 
centrifugal force. A necessary condition for self-confinement is that 
the total value $T_{\phi \phi} <0$, which reads:
\begin{equation}
    (e+P)\Gamma^{2} \Omega^{2}  \bar r^{4} < \frac{b^{2}}{4\pi}
    \label{eq:CONF}
\end{equation}
The toroidal field, and thus the current, decreases with mass density 
along the jet. Because $l \simeq l_*$ for $m^2 \gg 1$, the collimation requirement 
becomes asymptotically
\begin{equation}
    b^2/4\pi > \rho l_{*}^{2}
\label{eq:CONDF}
\end{equation}
Although self-collimation can be realized in non-relativistic jets, numerical works have shown a lack of collimation of relativistic 
jets (\cite{SB} \cite{BT}). This suggests that the tension is unable 
to balance the internal pressure in relativistic jets. The difference 
between relativistic jets and non-relativistic ones comes from the 
importance of the electric field in the former case. 
Indeed $B_{\phi}^{2} = 
B_{p}^{2}(\Omega_{*}-\Omega)^{2}\bar r^{2}/v_{p}^{2}$ and 
$E^{2}=B_{p}^{2}\Omega_{*}^{2}\bar r^{2}/c^{2}$. This clearly shows that, for 
$v_{p}=c$, the toroidal magnetic field is a little smaller than the 
electric field, and, thus, its tension is unable to balance the 
pressure due to the electric field.
\begin{equation}
\label{ }
\frac{B_{\phi}^2-E^2}{B_p^2} = -2\frac{l_*\Omega_*}{\Gamma c^2} \rightarrow -2
\end{equation}
On the contrary, with non-relativistic motions, cylindrical 
self-collimation is possible, as rigorously proved by Heyvaerts and Norman \cite{HN}. 
Even recollimation is possible after the flow has passed the fast magnetosonic 
surface (\cite{PP}, \cite{F}), thus assuming a tulip shape.

These statements can be set more accurately. Suppose that a cylindrical collimation is realised
with a necessarly finite amount of current intensity $I(\bar r) \rightarrow I_{tot}$ at large distance from axis.
Let $P' \equiv P+\frac{B_p^2+ E^2}{8\pi}$; then the confinement is such that the toroidal field contribution balances the gradient of $P'$ and the centrifugal effect:
\begin{equation}
\label{ }
\frac{\partial P'}{\partial \bar r} -w\rho \Gamma^2 \Omega^2 \bar r = -\frac{1}{8\pi \bar r^2} \frac{\partial}{\partial \bar r}(\bar rB_{\phi})^2 \ .
\end{equation}
First, suppose that $E^2 \ll B^2$, as it is the case in non-relativistic jets; then one derives from the confinement equation the required current:
\begin{equation}
\label{ }
\frac{I^2(\bar r)}{2\pi} = \int_0^{\bar r} (2rP'+ w \rho\Gamma^2 \Omega^2 r^3)dr - \bar r^2P'(\bar r)
\end{equation}
$B_p^2$, $P'$ and $\rho$ are supposed to radially decrease so that the integral converges, and thus the existence of a jet radius means that at larger radii the current saturates at a finite value.
\begin{equation}
\label{ }
\frac{I^2_{tot}}{2\pi} = \int_0^{\infty} (2rP'+ w\rho\Gamma^2 \Omega^2 r^3)dr \ .
\end{equation}
A jet radius $\bar r_j$ can therefore be defined such that $I^2_{\infty} = 4\pi \bar r_j^2 P'(0)$.

Now, consider a relativistic jet developing an electric field such that $E^2 > B_{\phi}^2$. Take the estimate $E^2 = B_{\phi}^2 + 2B_p^2$ (in fact, the poloidal contribution is not important, as will be seen below). Then the confinement equation is changed into:
\begin{equation}
\label{ }
\frac{\partial P"}{\partial r} -w\rho \Gamma^2 \Omega^2 \bar r = -\frac{1}{\pi \bar r^2} (\frac{\partial}{\partial \bar r}I^2- \frac{I^2}{\bar r}) \ .
\end{equation}
The modified pressure term $P"$ now contains $3B_p^2/8\pi$ instead of $B_p^2/8\pi$.
The solution is such that
\begin{equation}
\label{ }
\frac{I^2(\bar r)}{\pi} =  \bar r \int_0^{\bar r}(P"+ w\rho\Gamma^2 \Omega^2 r^2)dr - \bar r^2 P"(\bar r) \ .
\end{equation}
Since collimation is assumed, the integral must converge and $\bar r^2 P"(\bar r) \rightarrow 0$. But then
$I$ should be proportional to $\sqrt{\bar r}$, which is in contradiction with a finite confining current, since the existence of jet radius $\bar r_j$ would be characterized by a convergence of the current for $\bar r \gg \bar r_j$. Therefore it is impossible for a relativistic jet to undergo self-collimation. As anticipated above, the modification of the pressure term ($P"$ instead of $P'$) does not make a qualitative difference and, even with a lower estimate of the electric field, $E^2 = B_{\phi}^2$, collimation is impossible.

To achieve self-collimation, the jet must be composed of a relativistic flow (hereafter R-jet) close to the axis embedded by a non-relativistic one (hereafter NR-jet) that insures collimation. The relativistic is confined in a radius $r_*$ and the non-relativistic one in a radius $r_j$ and they are related to the currents through the following relations:
\begin{equation}
\label{ }
\frac{I^2(r_*)}{\pi} =  r_* \int_0^{r_*}(P"+ w\rho\Gamma^2 \Omega^2 r^2)dr - r_*^2 P"(r_*) \ ,
\end{equation}
and
\begin{equation}
\label{ }
\frac{I^2_{tot}}{2\pi} = \frac{I^2(r_*)}{2\pi}+ \int_{r_*}^{\infty} (2rP'+ w\rho \Omega^2 r^3)dr +r_*^2P'(r_*) \ .
\end{equation}
By inserting profiles for the pressure and the mass density in these integrals, one gets a relation between the radius and the current. For smooth profiles in the relativistic jet, one roughly gets the relation
\begin{equation}
\label{ }
\pi (e_*+P_*)l_*^2(r_*) \simeq I^2(r_*) \ .
\end{equation}
And for profiles like $(\frac{r_j^2+r_*^2}{r_j^2+r^2})^2$ for $P'$ and $\rho$ in the non-relativistic jet, one gets, for $r_j \gg r_*$
\begin{equation}
\label{ }
I^2_{tot} \simeq I_*^2(r_*) + 2\pi (P'(r_*)+{1\over 2} \rho(r_*) \Omega_*^2(r_*)r_*^2)r_j^2 \ .
\end{equation}
Since the angular momentum $l_*$ is related to the initial current through $l_* \simeq -I_0/2\pi \kappa$, and reminding that $4\pi \kappa^2 = \rho_A/w_A$, one has
\begin{equation}
\label{ }
\frac{I^2(r_*)}{I_0^2} = \frac{e_*+P_*}{e_A+P_A}w_A^2\mid_{{\cal S}_*} \ ,
\end{equation}
where ${\cal S}_*$ labels the interface between the R-jet and the NR-jet. This stresses one more time the necessary decrease of the electric current in jet generation. Another important remark is that the power of the NR-jet needs to be significant in order to confine the R-jet, since the power is proportional to the current, at least in the launching stage.

\section{Jets from an accretion disk}

The accretion power is closely related to the process of angular 
momentum extraction from the disk. Powerful jets can be generated from 
the accretion disk provided that they play a role in the angular 
momentum transfer. There are four possibilities for transfering 
angular momentum of a disk. (In this section, the cylindrical coordinates $(r, \phi, z)$
are used)
\begin{itemize}
    \item  
The usual viscous transfer:
$$
\tau_{\phi r} = \eta_{vis} r\frac{\partial \Omega}{\partial r}
$$
which is very unefficient to account for the observed disk luminosity.

    \item  The transfer by the MHD turbulence stress, as generated by 
    MRI for instance:
    $$
    \tau_{\phi r} = <B'_{\phi} B'_{r}/4\pi -\rho u'_{\phi} u'_{r}>
    $$

    \item  Instead of being generated by an MHD turbulence, this 
    stress can be generated by a large amplitude wave with $m=1$ excited by the spiral wave 
    instability when the magnetic field is above the equipartition value (\cite{TP} \cite{VT}).

    \item  Instead of transfering radially, a static magnetic field 
    with opened lines in a bipolar configuration exerts a stress that 
    carries the angular momentum vertically:
    $$
    \tau_{\phi z} = B_{\phi} B_{z}/4\pi
    $$
\end{itemize}
This last possibility deserves a specific investigation, for this is 
precisely what jets are doing.

\subsection{Angular momentum transfer by the jet}\label{subsec.AMT}

It is straightforward to compare the torque due to the static
magnetic field, that has both a significant vertical component and toroidal component, to the one 
exerted by MHD turbulence in its most 
efficient regime where $\tau_{\phi r} \sim P_{m}$ (see the section devoted to the magneto-rotational instability). The torque per unit 
volume is $r \nabla_{j} \tau_{\phi}^j$. Thus the ratio of the 
turbulent torque over the static field torque is
\begin{equation}
    \frac{turb \, \, torque}{field \, \, torque} \sim \frac{hB_{z}}{rB_{\phi}}
    \label{eq:RAPTOR}
\end{equation}
It is easy to get $\vert B_{\phi} \vert > \frac{h}{r}B_{z}$. Thus the vertical extraction
of angular momentum by the jets is easy compared to the radial 
transfer by turbulence (\cite{PN} \cite{PP}). Actually, the main problem will be to avoid a 
too strong generation of $B_{\phi}$ by differential rotation in the accretion 
disk by an appropriate resistivity (this is likely an effective 
resistivity that accounts for reconnections in the disk) (\cite{FP}).

The transfer of angular momentum relates the accretion rate to the 
torque (bearing in mind that $b(r,-z)=-b(r,z)$):
\begin{equation}
    \dot M_{a}(l-l_{i}) = -\int_{r_{i}}^r b^{+}B_{z}rdr -2\pi r^{2} \int \tau_{\phi r} dz
    \label{eq:ACTOR}
\end{equation}
The Shakura-Sunyaev prescription consists in assuming $\tau_{\phi r} = - \alpha_v \rho \Omega^2h^2$ or a viscosity of the form $\nu_v = \alpha_v \Omega h^2$, and it is expected that the turbulence allows to reach a value of $\alpha_v$ as high as $0.1$. Similarly a turbulent magnetic diffusivity $\nu_m \equiv \eta/4\pi$ can be produced such that $\nu_m = \alpha_m \Omega h^2$, with $\alpha_m \sim \alpha_v$. The angular momentum extraction by the static field can be characterized by a parameter $\mu$ such that $B_{\phi}^+B_z/4\pi = - \mu \rho \Omega^2h^2$; $\mu$ is supposed smaller than unity, as well, for a reason that will be explained later on.
The accretion rate that stems from both kinds of angular momentum transfer is thus
\begin{equation}
\label{ }
\dot M_a \simeq 4\pi(\alpha_v h + \mu r)\rho \Omega h^2 \ .
\end{equation}
A magnetic Reynolds number ${\cal R}_m$ can be defined as ${\cal R}_m = ru_0/\nu_m$, where $u_0$ is the accretion velocity and its value is ${\cal R}_m \simeq {\alpha_v \varepsilon + \mu \over \alpha_m \varepsilon}$ where $\varepsilon$ is an appropriate average of $h/r$.

Assuming that the magnetic stress extracts a fraction $\chi_{j}$ of 
the angular momentum and the turbulent stress a fraction 
$\chi_{d}$ ($\chi_{j}+\chi_{d}=1$), then 
\begin{equation}
    rb^{+}B_{z} = -\chi_{j} \dot M_{a} \frac{\partial l}{\partial r}
    \label{eq:MAGEX}
\end{equation}
whereas
\begin{equation}
    \frac{\partial}{\partial r}(2\pi r^2 \int \tau_{\phi r} dz) = 
    -\chi_{d} \dot M_{a} \frac{\partial l}{\partial r}
    \label{eq:TURBEX}
\end{equation}
The power budget is therefore, as follows; for each jet:
\begin{equation}
    P_{j} = \frac{\chi_{j}}{2}\frac{1}{2}\frac{GM\dot M_{a}}{r_{i}}
    \label{eq:PJET}
\end{equation}
and for radiation:
\begin{equation}
    L = \chi_{d}\frac{1}{2}\frac{GM\dot M_{a}}{r_{i}}
    \label{eq:LUM}
\end{equation}
In fact the accretion power is shared in several contributions, that are simply expressed in terms of the Reynolds number ${\cal R}_m$ and the magnetic stress parameter $\mu$. When the magnetic Prantdl number is supposed to be 1, for simplicity,  we have :
\begin{itemize}
  \item 
 a Joule heating contribution
\begin{equation}
\label{ }
P_{Joule} \sim \frac{\mu}{{\cal R}_m}P_{acc} \ ,
\end{equation}
\item 
 a viscous heating contribution
\begin{equation}
\label{ }
P_{vis} \sim \frac{1}{{\cal R}_m}P_{acc} \ ,
\end{equation}
 \item 
therefore a luminosity
\begin{equation}
\label{ }
L \sim \frac{1}{{\cal R}_m}P_{acc} \ ,
\end{equation}
 \item 
and a power for each jet
\begin{equation}
\label{ }
P_{jet} \sim {1\over 2} (1- \frac{1}{{\cal R}_m}) P_{acc} \ ,
\end{equation}
 \end{itemize}
where the Reynolds number
\begin{equation}
\label{ }
{\cal R}_m \sim 1 + \frac{\mu}{\varepsilon} \ .
\end{equation}
Thus disregarding the small contribution of the Joule heating, one has $\chi_d \sim 1/{\cal R}_m$ and $\chi_j \sim 1-1/{\cal R}_m$.

There is a relation between the accretion rate and the ejection rate derived from the conservation of the angular momentum flux in the jet, indeed the angular momentum flux in the jet, $l_*d\dot M_j$, comes from the fraction of the fraction of angular momentum extracted from the disk by the magnetic stress :
$$
d \dot M_{j} = \frac{\chi_{j}}{2} \dot M_{a}\frac{dl}{l_{*}} \ ;
$$
which shows that the widening of the magnetic surfaces allows to transfer the angular momentum
with a rather weak ejection rate, provided that the AlfvŽn radius (the "lever arm") be large enough, 
since, roughly, $\dot M_j/\dot M_a \sim \chi_j r_0^2/r_A^2$.

The first solutions derived by Ferreira \& Pelletier (\cite{FP}) were obtained even with the extreme assumption of no other angular momentum transfer than through the magnetic stress, which corresponds to $\mu \gg \varepsilon$. The conditions for this accretion-ejection process to work are quite stringent but sensible. They have been made more flexible by introducing also an effective viscosity and an additional heating at the surface of the disk (\cite{CF1} \cite{CF2}) and checked by direct numerical simulations (\cite{CK}). It turns out that without this additional heating, $\chi_{d} \ll 1$, most of the accretion power goes in the jet, whereas the additional heating at the surface makes the mass outflux 
easier and allows solution with sizeable conversion coefficients $0.1 < \chi_d \sim \chi_j < 1$.

Such a jet that takes its power from extracting a significant fraction of the disk angular momentum has specific properties that have been investigated in( \cite{PP}) and allow observational tests (\cite{PR}).

\subsection{Power and asymptotic motion}\label{subsec.PAM}

There is a natural scaling with the mass of the central Black Hole for 
the power and the mass flux of the jets launched by accretion disk. 
The mass ejection rate is a fraction $\xi$ of the accretion rate $\dot M_j = \xi \dot M_a$; this 
ratio depends on the details of the transport physics in the accretion 
disk but not of the Black Hole mass. The averaged asymptotic bulk motion is 
such that
\begin{equation}
    \Gamma -1 = \frac{P_{j}}{\dot M_{j}c^{2}}= 
    \frac{\chi_{j}}{4\xi}\frac{r_{g}}{r_{lso}}
    \label{eq:ASYM}
\end{equation}
The bulk motion is independent (or weakly dependent) of the Black 
Hole mass. This is probably the reason why similar motions are observed in 
quasars and micro-quasars. This argument favors the model of jet 
generation by the accretion disk. However, because $\Gamma -1 \simeq 2 \times 
10^{-2} \xi^{-1}$, $\xi \sim 10^{-3}$ is required to get a
relativistic motion with $\Gamma = 10$; which is difficult to obtain from an accretion 
disk (\cite{FP}); higher values are commonly obtained in numerical simulations.
However relativistic motions are not ruled out in the innermost region and could be obtained from a 
weakly radiative torus. These estimates must be linked with the results obtained with Bernoulli equation (see section (\ref{sec.RNRM})). This relates the angular momentum invariant with the ration $\xi$ since $\Gamma_{\infty}-1 \simeq \frac{\Omega_*l_*}{c^2} \simeq 2\times 10^{-2} \xi^{-1}$.

\subsection{Conditions for ejection}\label{subsec.CE}

The accretion flow bends the field lines and the generation of a strong 
toroidal field by differential rotation must be prevented by the 
appropriate resistivity. Assume a resistivity profile $\nu_{m} = 
\nu_{m0}(1-\frac{z^{2}}{h^{2}})$ with a kind of Sakura-Sunyaev 
prescription $\nu_{m0} = \alpha_{m} \Omega h^{2}$, then the vertical 
variations of the 
field components can be expanded in Legendre polynomials. In this 
idealised description, the matching between resistive MHD inside the 
disk and ideal MHD in the jet has to be set at the disk edge 
$z=h(r)$. The value of $\alpha_{m}$ directly controls the relation 
between toroidal and radial components at the surface of the disk, 
$B_{\phi}^{+}$ and $B_{r}^{+}$: 
$$
B_{\phi}^{+} = -\frac{B_{r}^{+}}{\lambda \alpha_{m}} \ .
$$
The parameter $\lambda$ is a pure number that depends on the geometry 
of the field lines inside the disk, which is the solution of a kind of 
nonlinear eigen-value problem (\cite{PF}). For instance, the less "wavy" solution, 
where $B_{\phi}$ varies almost linearly with $z$ across the disk,
leads to the lowest value of $\lambda$ (ground state), which is close to unity. 
However, because $\alpha_m$ is expected to be significantly smaller 
than unity and $B_{\phi}$ not significantly stronger than the other 
field component, in order to prevent a strong squeezing of the disk, 
a higher value of $\lambda \sim 10$ is better, which is obtained by 
the second eigen-state, where $B_{\phi}$ has a cubic like variation 
across the disk. The poloidal field is quartic rather than quadratic 
in the vicinity of the equatorial plane.

It is expected that the field lines bend above the disk in order to 
fulfill the Blandford \& Payne (\cite{BP}) condition for launching in ideal MHD 
regime, namely a bending of more than $30^0$. Moreover it is 
expected that the field does not squeeze the disk significantly: 
$P_{m} < P$. In fact, it is possible to get $B_{\phi}^{+} \sim 
B_{r}^{+} \sim B_{z}^{+}$ together with $\alpha_{m} \sim 0.1$ (\cite{CF2}).

In the neighbourhood of the disk surface, matter is expelled provided 
that:
\begin{equation}
    \rho \frac{\partial \Phi_{eff}}{\partial s}+ \frac{\partial P}{\partial s}
    +\frac{1}{8\pi r^{2}}\frac{\partial b^{2}}{\partial s} < 0 \ ,
    \label{eq:EXPUL}
\end{equation}
where $s$ is a coordinate along the poloidal field line that vanishes 
at the foot of the line. Expansion of 
the effective potential for small $s$ leads to $\frac{\partial 
\Phi_{eff}}{\partial s} = \Omega_{0}^{2}(3\tan^{2}\psi -1)z \cos 
\psi$, where $\psi$ is the angle of the tangent to the field line with respect to the vertical direction. 
Therefore, if the field structure is almost force free 
($b(a)$) and matter cold, the condition for an outflow is that the 
field line bends by an angle larger than $\pi/6$. The effective force 
vanishes at the point where $\psi = \pi/6$. Consider now, finite 
temperature and non force free effect. If at this point, MHD is 
already ideal, the current is decreasing and acts like the pressure 
gradient to expell matter. But if MHD is still resistive (inside the 
disk), the toroidal field squeezes the disk and tends to prevent 
matter escape. Thus the pressure gradient must overcome the squeezing 
effect; the condition is roughly $P > (B_{\phi}^{+2}+B_r^{+2})/8\pi$ or 
simply $P > P_m$. This is the reason 
why there is a compromise for the intensity of the magnetic field: it 
must be high enough to extract the angular momentum, but the 
generation of the toroidal field must remain moderate to avoid the 
disk squeezing that prevent matter escape. This compromise is a 
magnetic field intensity of the order of equipartition value. Clearly 
the previous condition indicates that a hot corona helps matter to 
escape. When an additional coronal heating is at work (\cite{CF2}), the mass 
ejection rate is related to the coronal temperature.

Although the first semi-analytical solutions (\cite{FP}) to the accretion-ejection problem were derived with a self-similar assumption, that cannot be valid at both edges of the accretion disk, they revealed the main constraints for an accretion disk to launch jets. The main features of these analysis were confirmed by direct numerical solutions (\cite{CK}).
Now to get fair simulations of the asymptotic behaviour of jets, 3D-simulations are needed to understand how the development of 3D-instabilities does not destroy the jet (\cite{OP}, \cite{BK}).

\subsection{Resistivity problem}\label{subsec.RP}

In a SAD, an effective viscosity is invoked to get the expected accretion rate to account for the luminosity 
of the disk. In AGNs for instance, the viscosity coefficient $\alpha_v$ of the Shakura-Sunyaev prescription (e.g. $\tau_{\phi r} = \alpha_v \rho\Omega^2 h^2$) is expected to be on the order of $0.1$ and supposed to stem from a turbulent state of the accretion flow. If the angular momentum is vertically extracted from the disk by the magnetic field, as explained previously, then there is no viscosity problem. But there is a resistivity problem, because a quasi stationary accretion flow cannot be set up if matter does not easily cross the magnetic surfaces. A magnetic diffusivity is required as expressed by the prescription $\nu_m = \alpha_m \Omega h^2$ with $\alpha_m \sim 0.1$ as well. In fact, a fully developed state of MHD turbulence leads to turbulent magnetic diffusivity comparable to the turbulent viscosity. But as shown before, even if such a turbulent transport is achieved, the vertical transport of angular momentum by the jet can easily dominate the accretion dynamics.

Knowing the cause of the excitation of turbulence in an accretion disk is an important issue that receives crucial contribution during the last decade. 

\subsection{Magneto-Rotational Instability}\label{subsec.MRI}

The magneto-rotational instability has been considered as a good driver of the turbulence in the disk, with the capability of transferring the angular momentum radially (\cite{VE}, \cite{CH} , \cite{BH}).
A brief presentation of the Magneto-Rotational Instability (MRI) will be done, using the formalism previously introduced in section (3) and subsection (4.1), and will be limited to the following assumptions for the sake of simplicity: the analysis is done in a thin equatorial disk far enough from the Kerr Black Hole where $\omega \simeq 2a_s c r_g^2/r^3$ and $H_{r\phi} \simeq \omega$ (see subsection \ref{subsec.SPLIT}), with a vertical mean field $B_0(r)$, with incompressible perturbations. The main purpose is to look at whether both MRI-instability and jet launching are compatible in an accretion disk.
The radial equilibrium for ZAMOs is such that the centrifugal force modified by the gravito-magnetic effect balances the gravitational attraction $g_r = -c^2 {\partial \over \partial r} \log \alpha$ with $\alpha^2 \simeq 1-2r_g/r$; which reads
\begin{equation}
\label{ }
\Omega^2 + \omega \Omega = \Omega_0^2 \equiv -c^2 {1\over r}{\partial \over \partial r} \log \alpha
\end{equation}
Consider a small radial displacement $\xi$ such as $\alpha u_r = {\partial \xi \over \partial t}$. Since the magnetic surfaces are frozen in the flow as expressed by eq.(\ref{eq:EVA}), the linearised version reads : 
$$
\tilde a \equiv \delta a+ \xi 
\frac{\partial a}{\partial r} = 0
$$
The evolution of the radial displacement (Alfv\'en wave modified by centrifugal force) is governed by:
$$
\frac{\partial^{2}}{\partial t^{2}}\xi -  
V_{A}^{2}\frac{\partial^2}{\partial z^2}\xi -r(2\Omega +\omega)\delta \Omega = 0
$$
The angular velocity undergoes Lagrangian variations $\tilde \Omega$ such that: 
$$
\tilde \Omega \equiv \delta \Omega + \xi 
\frac{\partial \Omega}{\partial r} \ .
$$
Note that the Lagrangian variation of the specific angular momentum $\tilde l = \delta l + \xi \frac{\partial l}{\partial r}$ differs from $\tilde \Omega r^2$ by the Coriolis contribution $2\Omega r\xi$. The equation governing the radial displacement can be rewritten as:
$$
\frac{\partial^{2}}{\partial t^{2}}\xi - 
V_{A}^{2}\frac{\partial^2}{\partial z^2}\xi +r(2\Omega + \omega)\frac{\partial 
\Omega}{\partial r} \xi = r(2\Omega + \omega) \tilde \Omega
$$
This form of the equation already shows that the angular velocity shear is the motor of the instability.
The coupling with the Lagrangian variation of the angular velocity will introduce more inertia due to Coriolis 
force. A second equation coupling $\tilde \Omega$ with $\xi$ is obtained by the transport of angular momentum and the induction equation, by eliminating the toroidal field $b$ between them:

\begin{equation}
\label{EVLL}
\frac{\partial \tilde l}{\partial t} = \frac{B_0}{4\pi \rho}\frac{\partial b}{\partial z}
\end{equation}

\begin{equation}
\label{EVBL}
\frac{\partial b}{\partial t} = B_0(\frac{\partial \tilde \Omega r^2}{\partial z} + 
r^2\frac{\partial \omega}{\partial r} \frac{\partial \xi}{\partial z}) \ .
\end{equation}
The second equation describing the MRI instability is thus obtained:
\begin{equation}
\label{ }
\frac{\partial^{2}}{\partial t^{2}} \tilde \Omega -V_A^2 \frac{\partial^2}{\partial z^2} \tilde \Omega = -2\frac{\Omega}{r} \frac{\partial^2}{\partial t^2} \xi +V_A^2 \frac{\partial \omega}{\partial r} \frac{\partial^2 \xi}{\partial z^2} \ .
\end{equation}
Looking for eigen-modes and defining the appropriate vertical wave numbers $k_n$, which has a minimum value $\sim \pi/h$, the following condition for the development of the MRI instability is derived:
$$
k_{n}^{2}V_{A}^{2} + (2\Omega + \omega) r\frac{\partial}{\partial r} (\Omega + \omega) <0 \ .
$$
The condition can be rewritten in the form :
\begin{equation}
\label{ }
k_{n}^{2}V_{A}^{2} < 3\Omega_0^2 (1 + \frac{r_g}{r} + 2a_s (\frac{r_g}{r})^{3/2} + ...) \ .
\end{equation}
This result is close to Charles Gammie's result (\cite{CG}), it differs by the sign of the Black Hole rotation contribution in $a$, which increases the instability in this result. The result is obtained by expanding in 
$r_g/r$, however it seems that the instability is stronger at shorter distance.
The interest of the exercise is two-fold. First, disregarding the gravito-magnetic contribution, it shows that there is a threshold condition for the growth of the instability in an accretion disk, because of the minimum value of the wave number. There is an
instability only for $B<B_{c}$ such that $B_{c}^{2}/8\pi \simeq P$. This condition is more or less in contradiction with the condition for jet launching. In other words, the interval of magnetic field intensities in which both the MRI develops turbulence and jets are launched is narrow. Second, the gravito-magnetic contribution in the vicinity of a Kerr Black Hole enlarges this interval and strengthens the instability together with the Schwarzschild contribution in $r_g/r$. Anyway another instability would be welcome...
A detailed analysis of the MRI instability in the vicinity of a Kerr Black Hole will be found in a forthcoming paper (\cite{PF}).

The MRI has a maximum growth rate of order $\Omega$ for $k_{max} = \Omega/V_A$, if larger than  $\pi/2h$, and  excites 3D-turbulence at scale $h$ and leads to a stress $\tau_{\phi r} \sim P_m$. The resistivity is such that the effective magnetic diffusivity remains below but on the order of the value which would remove the frozen in condition for the mean field;
 $$
\nu_m \sim \Omega/k_{max}^2 \sim V_A^2/\Omega \sim (P_m/P) \Omega h^2 \ ,
$$
which is the expected height, but is valid for $P_m < P$, and corresponds to a Shakura-Sunyaev prescription with $\alpha_v \sim \alpha_m \sim P_m/P < 1$.

\section{Two-flow model for AGNs and Micro-Quasars}\label{sec.}

The two-flow model consists in combining a relativistic flow along the 
axis and a non-relativistic or mildly relativistic flow coming from 
the accretion disk. When this model has been proposed in the eighties (mostly with a description of the interaction of the cold relativistic beam with an ambient medium) (\cite{PNICE}), the arguments were rather poor: the phenomenology of hot spots do not require a relativistic flow, whereas relativistic sporadic ejections are observed only at short distance from the source; the large scale jet does not seem to  loose significant energy during its travel, and non-relativistic shock acceleration satisfactorily explains the hot spot synchrotron spectra. So the large scale NR-jets and the short scale R-jets could be generated differently.  First astrophysical applications of the model was published in 1989 (\cite{SPA} \cite{PERO} \cite{PESO}). But more arguments have been proposed later (\cite{HP}, \cite{HPR}) in relation with X-ray and gamma-ray emissions. The interest of the two-flow model relies on the following main 
arguments:

i) The relativistic jet alone would radiatively cools rapidly and be
relaxed by Compton drag (\cite{PS}). As we previously saw, Blandford \& Znajek mechanism works with the bunch of magnetic surfaces that thread the Black Hole ergosphere thanks to falling matter carrying negative total angular momentum (mostly due to the toroidal magnetic field). That matter is necessarily created above the polar cap, and cannot be anything else than pairs. Above the stagnation surface the outflow is made of pairs. They are unavoidably experiencing the Compton drag. People claim that there would not be any Compton drag problem with protons. But protons can come in the relativistic jet only by diffusion from its environment (accretion disk or MHD NR-jet) and then be entrained.
This amount of protons is very tiny up to a large distance beyond which ambient protons fill up the relativistic flow.

ii) The relativistic jet does not self-collimate. The non-relativistic flow, which self-collimates, can confine it and also manages an ecological niche for it, because of the centrifugal barrier inherently associated with the MHD acceleration of the NR-jet. A simulation of a collimating shock between the R-jet and the NR-jet has been done (\cite{TB}).

iii) The extraction of rotation energy from a spinning Black Hole 
does not generate a power strong enough to account for the large power 
of FR2 jets. Jets powered by accretion are more powerful, but are 
mildly relativistic.

iv) The R-flow is likely in the form of flaring clouds, rich in $e^+-e^-$-pairs, channelled by the NR-jet, and involving a smaller power. It probably does not travel beyond kpc distance from its source.

\begin{figure}[h]     
\begin{center}     
\caption{The two-flow model (from L. SaugŽ thesis manuscript). The non-relativistic MHD jet, launched by the accretion disk, manages a centrifugal barrier along the axis where pair clouds can be generated by photon-photon interactions. In the anisotropic radiation field emanating from the accretion disk, these pair clouds suffer a Compton drag or propulsion depending on their velocity and energy distribution, and they are channeled by the magnetic surfaces. They are maintained with a hot relativistic distribution by in situ acceleration through the MHD jet turbulence.}     
\label{fig2}     
\end{center}     
\end{figure}     

A new version of the two-flow model has been developed since 1991, where the R-jet is a hot pair plasma, energised by turbulence from the NR-jet (\cite{HP}, \cite{HPR}) and experience a Compton Rocket effect. Satisfactory fits of the spectra of 3C273 and 3C279 has been obtained, with a natural break at a few MeV (\cite{MHP}); the stratification plays a major role in the analysis, which deeply differs from homogeneous or slab models. Later, a study of the micro-instabilities of such a R-jet have been investigated (\cite{MPH}); these instabilities differ from the one investigated in the previous version, by the fact that they are triggered by the streaming of the cold ambient medium in the hot R-jet, instead of considering the R-jet as a cold beam exciting waves by pervading the ambient medium. This is an interesting circumstance of plasma turbulence excitation, because the streaming instability is driven by the radiation force and quasi-linear theory is not sufficient to describe the stationary state; non-linear mode couplings are necessarily taken into account, and, in particular, allow to estimate the heating of the pair plasma.

\subsection{Compton Rocket}\label{subsec.CR}

Following O'Dell (\cite{ODE}), the anisotropic radiation field of soft photons, emanating from the accretion disk, or from the torus, generates a Compton force on electrons and positrons that can be pushing or dragging depending on the velocity of the flow.
The component $F_0 \equiv \vec F.\vec v$ of the force depends on the three Eddington momenta of the radiation field, on the velocity $\beta$ of the radiating electron and its pitch angle cosine $\mu$; it is such that:
\begin{equation}
\label{FC}
\frac{d\epsilon}{dt} = F_0 = -2\pi \sigma_{T}\gamma^{2}\beta 
[\beta(3J-K)-2(1+\beta^{2})H\mu + \beta(3K-J)\mu^{2}] \ ,
\end{equation}
where $J$, $H$ and $K$ are the Eddington coefficients, momenta of the 
intensity function :
$$
J = \frac{1}{2}\int \int I(\nu, \vec n) d\nu d\mu_s \ ,
$$
$$
H = \frac{1}{2}\int \int I(\nu, \vec n) \mu_s d\nu d\mu_s \ ,
$$
$$
K = \frac{1}{2}\int \int I(\nu, \vec n) \mu_s^{2}d\nu d\mu_s \ .
$$
\bigskip
These momenta have been calculated for a SAD and
there exists an equilibrium velocity $\beta_{eq}$ that makes the bracket in $F_0$ to vanish, such that $\frac{d\epsilon}{dt} = 0$ (\cite{HP}, \cite{MHP}, \cite{RH}). 
For $\beta < \beta_{eq}$, the radiation force accelerates, because the head on collisions of the electron (positrons) with photons favour a forward momentum in statistical balance, whereas it decelerates
for $\beta > \beta_{eq}$ (Compton drag), because electrons (positrons), advancing at relativistic motions, suffer more recoil effect in the statistical balance. For a SAD, $\Gamma \simeq (\frac{z}{r_{i}})^{1/4}$ 
when $r_{i} \ll z \ll r_{e}$, $r_i$ and $r_e$ respectively being  the internal and external radii of the SAD, and $\Gamma \simeq (\frac{z}{r_{i}}^{1/4} r_{e}^{3/4})$ when 
$z \gg r_{e}$. A detailed analysis of such a  propulsion has been done by Renaud and Henri (\cite{RH}) for AGNs and micro-quasars. The results for micro-quasars cannot simply be obtained by a black hole mass re-scaling because the Klein-Nishina regime of the Compton effect is more important than in the case of AGNs (see fig(3)). Clearly, the Compton rocket is less efficient when the Klein-Nishina regime becomes significant. Moderate bulk Lorentz factors were obtained, but they are sufficient to account for the observations.
\begin{figure}[h]     
\begin{center}     
\caption{The jet bulk Lorentz factors (from \cite{RH}). The terminal bulk Lorentz is obtained as a function of $\gamma_{max}$, which is more or less the mean energy of the pairs in the co-moving frame, for different values of the distribution index $s$. The two top panels corresponds to a $10^9M_{\odot}$-black hole with $r_e = 3\times 10^3 r_g$, $L= L_{Edd}$ (left) and $L= 0.1L_{Edd}$ (right). The two bottom panels correspond to a $5M_{\odot}$-black hole with $r_e = 3\times 10^3 r_g$, $L= L_{Edd}$ (left) and $L= 0.1L_{Edd}$ (right).}     
\label{fig3}     
\end{center}     
\end{figure}     
In fact, a recent study by Henri and SaugŽ shows that the blazar bulk Lorentz factor should not be very large, larger than 2 but significantly smaller than 10 (\cite{SH}), contrarily to the estimate derived from homogeneous models. Indeed homogeneous models provides Doppler factors larger than 30 when the correction of the Infrared background absorption is done (even by lowering the doubtful point at $60 \mu m$.
These high estimates are in disagreement with the reasonable assumption that BL-Lac objects are normal FR1 jets seen at narrow angle, which leads to statistical estimates of bulk Lorentz factors between 3 and 5. Stratified models where the X-ray synchrotron emission is largely due to other electrons than those emitting the inverse Compton emission can fit the compound spectra with these low values of the Lorentz factor. However they require local quasi mono-energetic distribution functions rather than power law distributions and an opacity implying a significant pair creation (see \cite{PHGS}). These observational constraints revealing the importance of pair creation has a tremendous impact on the generation of the jets, as emphasized in several sections. The fact that the local distribution functions are likely quasi mono-energetic rather than of power law type implies that the acceleration process at work is more likely a kind of second order Fermi process or due to an electric field in reconnection sites.

\subsection{Variability}\label{subsec.VAR}

The physics involved in the two-flow model makes the relativistic jet intrinsically variable. If the turbulence heating of the pair plasma is maintained in a region where $\tau_{\gamma \gamma} > 1$, pairs create $\gamma$-rays by inverse Compton effect, that are converted into new pairs, which are accelerated and then create also $\gamma$-rays, and so forth. This is the pair creation catastrophe, which leads to a highly unstable situation. Its simulation shows various regimes of flaring very suggestive of the Blazar phenomenon, both qualitatively and quantitatively (amplitude and time scale) (\cite{SH}). These regimes of sporadic flaring depend on the power of the turbulence that maintains the in situ Fermi acceleration (of second order type). The level of the turbulence energy density varies with the energy mass density of pairs that absorb the power input: the increase of the pair energy-mass leads to more absorption of AlfvŽn waves energy and thus reduces the efficiency of the in situ acceleration.

\section{Discussion}\label{sec.DISCUT}

In these lectures, the main elements for studying the formation of jets in Black Hole environments have been presented together with the important opened questions that deserve continuing the theoretical effort for the achievement of this major topic of modern astrophysics. After having exposed the common knowledge in this field that has been developed during more than twenty years, some specific views about the way to solve some major difficulties has been proposed. Let summary the main problems encountered  for relativistic jet formation. 

First, the large scale AGN jets do not seem to be relativistic, especially FR1, and the energy budget of FR2 hot spots does not clearly imply a relativistic inflow, whereas their power is comparable to the accretion power, which means that they do not lose significant energy during their travel. Unsteady relativistic ejections are obvious at VLBI scales, suggesting a flaring activity. Note that the GRBs collimated winds, probably highly relativistic, are also very unsteady. 

Second, relativistic MHD flows have not the self-collimating property of non-relativistic ones. An external confinement is required but cannot be done by the intergalactic medium in the case of the well collimated FR2 jets.

Third, the power of the Blandford-Znajek mechanism does not seem sufficient to explain the powerful FR2 jets. But its power could be sufficient for generating the relativistic clouds at VLBI scales.

How Black Holes are generated with extreme rotation? Probably not by the accretion process if the Black Hole produces jets at the same time (\cite{HEN}). An equilibrium value of the spin parameter should be expected. Black Holes merging is probably necessary to get fast rotation. The fact that the Black Hole at the centre of Milky Way is fastly rotating is intriguing. Maybe because it was not producing jets...

Fourth, the generation of a relativistic flow is severely limited by the Compton drag exerted by the photon field emitted by the accretion disk. When Compton drag is responsible of its relaxation, within few tens of gravitational radii, the jet vanishes and gives rise to hard X-ray emission. This may happen in Seyfert galaxies, where no relativistic jets are seen.

For all these reasons, for AGNs and micro-quasars, we (myself and my collaborators) have proposed to circumvent these difficulties in the frame of a two-flow model that combines a non-relativistic or mildly relativistic jet launched by the accretion disk with a relativistic flaring ejection that is channelled by the sub-relativistic jet. An observational evidence of these combined flows is strongly suggested in Scorpius X1 (\cite{FOM1} \cite{FOM2}). Besides the confinement produced by the self-collimating NR-jet, the relations between the two flows are as follows. The NR-jet is more powerful and part of its MHD turbulence energises the R-jet plasma. This R-jet plasma is rich of $e^+e^-$ pairs; baryons are progressively entrained by the relativistic jet and then, after some travel of at least thousand gravitational radii and probably more, both jets mix together.
As explained in the text, the constant stochastic heating of particles of the relativistic jets strengthens the Compton force, both its drag effect and its rocket effect, and then significant Lorentz factors can be obtained, whereas, if not heated, the jet rapidly cools and slow down to a modest Lorentz factor. Moreover, this stochastic heating makes this ejection highly variable by stimulating the pair creation catastrophe.

The study of the two-flow model is not yet completed. It would be important to combine the Poynting flux propulsion, together with the effect of the Compton force in an anisotropic radiation field, taking account of the stochastic heating of pairs by turbulence and of a progressive baryonic pollution. Moreover, we are currently studying the changes of regime in the accretion-ejection flows, where a major part of the accretion disk sporadically becomes radiatively inefficient (see \cite{FHP}). During this stage, because the compactness drops, the pair creation vanishes, making the magnetisation higher, and the Compton drag weaker, which allows a relativistic jet generation by Blandford-Znajek effect. When, for some reason related to radiative properties of the disk, a SAD is again set up, pair creation and Compton drag become stronger again, which bridles the relativistic jet.

Among the issues that are currently investigated in this field, the magnetic coupling between the black hole and the disk (see for instance \cite{G}), and between the black hole, the disk and the jets, are difficult but crucial topics that deserve the development of a new generation of MHD numerical codes.

The generation of relativistic jets in the environment of Black Holes is not only a fascinating issue per se, but is also at the origin of high energy phenomena, that are currently at the forefront of the astroparticles physics.

\section{acknowlegment}

I feel grateful to Jacques Paul and his group of Saclay, especially Philippe Ferrando, Philippe Laurent and Christian Gouiff\`es, for their welcome and help at Cargse school in May 2003. I also thank Maxim Lyutikov for having invited me to participate in this interesting task and for his patience. My collaborators and friends, Jonathan Ferreira, Gilles Henri and Pierre-Yves Longaretti, are especially thank for their constant stimulating attitude and for sharing fruitful discussions. Ludovic SaugŽ is thank for his help in computing stuff and for the nice picture of the two-flow model he has made for his thesis and transmitted to me.


\begin{thebibliography}{}
\bibitem{ACGL}
Abramowicz M.A., Chen X.-M., Granath M., Lasota J.P., 1996, ApJ, 471, 762.
\bibitem{BH}
Balbus S., Hawley J., 1992, ApJ., 392, 662.
\bibitem{BK}
Baty H., Keppens R., 2002, ApJ, 580, 800.
\bibitem{BL}
Begelman M.C., Li Z-Y, 1994, ApJ 426, 269.
\bibitem{BEK}
Bekenstein J.D., 1973, Phys. Rev. D, 7, 2333.
\bibitem{BBR}
Begelman M.C., Blandford R.D., Rees M.J., 1984, Rev. Mo. Phys., 56, 255.
\bibitem{BEL}
Belloni,ÊT.; Klein-Wolt,ÊM.; MŽndez,ÊM.; vanÊderÊKlis,ÊM.; vanÊParadijs,ÊJ., 2000, A \& A, 355, 271.
\bibitem{VBP}
Beskin V.S., Pariev V.I., 1993, Phys. Uspekhi, 36, 529.
\bibitem{BR}
Beskin V.S., Rafikov R.R., 2000, MNRAS, 313, 433.
\bibitem{VS}
Beskin V.S., Zakamska N.L., Sol H., 2004, MNRAS, 347, 587.
\bibitem{BZ}
Blandford R.D. \& Znajek R.L., 1977, MNRAS 179, 433.
\bibitem{BP}
Blandford R.D., Payne D. G., 1982, MNRAS, 199, 883.
\bibitem{SB}
Bogovalov S., 2001, A \& A, 371, 1155.
\bibitem{BT}
Bogovalov S., Tsinganos,ÊK.: 2001, MNRAS, 325, 249.
\bibitem{CMa}
Camenzind M., 1997, "Active Galactic Nuclei", Lect. Notes Phys., (Springer-Verlag).
\bibitem{CMb}
Camenzind M., 2004, in "Accretion Disk, Jets and High Energy Phenomena in, Astrophysics", Les Houches session 78 (EDP Sciences and Springer).
\bibitem{CMc}
Camenzind M., 1986, A \& A, 156, 137.
\bibitem{CMd}
Camenzind M., 1986, A \& A, 162, 32.
\bibitem{CF1}
Casse F., Ferreira J.: 2000, A \& A, 353, 1115295, 807.
\bibitem{CF2}
Casse F., Ferreira J., 2000, A \& A, 361, 1178.
\bibitem{CK}
Casse F., Keppens R., 2004, ApJ, 601, 90.
\bibitem{CH}
Chandrasekhar S., (1960), {\it Proc. N.A.S.}, {\bf Vol. 46}, pp. 56
\bibitem{CM}
I.F. Mirabel, V. Dhawan, S. Chaty, L.F. Rodriguez, J.Marti, C.R. Robinson, J. Swank, and T.R. Geballe, 1997, A \& A,
\bibitem{CAL}
Chen X.-M., Abramowicz M., Lasota J.P., Narayan M., Yi I., 1995, ApJ, 443, L61.
\bibitem{FP}
Ferreira J., Pelletier G., 1995, A \& A, 295, 807.
\bibitem{F}
Ferreira J.: 1997, A \& A, 319, 340.
\bibitem{FHP}
Ferreira J., Henri G., Pelletier G., (in preparation).
\bibitem{FOM1} Fomalont, E.~B., 
Geldzahler, B.~J., \& Bradshaw, C.~F. 2001, ApJLet, 553, L27.
\bibitem{FOM2} Fomalont, E.~B., 
Geldzahler, B.~J., \& Bradshaw, C.~F. 2001, ApJ, 558, 283.
\bibitem{FN}
Frolov V.P., Novikov I.D., 1998, Black Hole Physics (Kluwer Academic Pub, Dordrecht)
\bibitem{G}
Gammie C., 1999, ApJ, 522, 141.
\bibitem{CG}
Gammie C., 2004, ApJ, 614, 309.
\bibitem{H}
Hawking S.W., 1974, Nature, 248, 30.
\bibitem{HEN}
Henri G., 2004, oral communication in "le remue-mŽninges des SHERPAS".
\bibitem{HP}
Henri G., Pelletier G., 1991, ApJ, 383, L7.
\bibitem{HPR}
Henri G., Pelletier G., Roland J., 1993, ApJ, (1993), {\it ApJ. Lett.}, {\bf Vol. 404}, pp. L41.
\bibitem{HN}
Heyvaerts J., Norman C.Z., (1989), {\it ApJ.}, {\bf Vol. 347}, pp. 1055.
\bibitem{KKS}
Katarzinsky K., Sol H., Kus A.: 2001, A \& A, 367, 809.
\bibitem{LBC}
Li Z.Y., Begelman M., Chiueh T., 1992, ApJ, 384, 567.
\bibitem{LOV}
Lovelace, R.V.E., Wang, J.C.L., Sulkanen, M.E. (1987), {\it ApJ}, {\bf 
Vol. 315}, pp. 504.
\bibitem{LB}
Lyutikov M., Blandford R.D., 2004, Astroph.
\bibitem{MHP}
Markowith A., Henri G., Pelletier G., 1995, MNRAS, 277, 681.
\bibitem{MPH}
Marcowith A., Pelletier G., Henri G., 1997, A \& A, 323, 271.
\bibitem{MRs}
Meszaros P., 2002, ARA \& A, 40, 137.
\bibitem{MRL}
Meszaros P., Laguna P., Rees M.J., 1993, ApJ, 415, 181.
\bibitem{M1}
Michel F.C., 1969, ApJ, 158, 727.
\bibitem{M2}
Michel F.C.,  1973, ApJ, 180, L133.
\bibitem{MR1}
Mirabel I.F. \& Rodriguez L.F., 1998, Nature, 392, 673. 
\bibitem{MR2}
Mirabel I.F. \& Rodriguez L.F.,1999, Annual Review of 
Astronomy and Astrophysics, Vol. 37, pp. 409-443.
\bibitem{MTW}
Misner C.W., Thorne K.S., Wheeler J.A., "Gravitation"
\bibitem{MOR}
Morganti,ÊR.; Robinson,ÊA.; Fosbury,ÊR.ÊA.ÊE.; diÊSeregoÊAlighieri,ÊS.; Tadhunter,ÊC.ÊN.; Malin,ÊD.ÊF., 1991, MNRAS, 249, 91.
\bibitem{NY}
Narayan R. \& Yi I., 1995, ApJ., 452, 710.
\bibitem{OP}
Ouyed R., Pudritz R.E., Stone J.M., 1997, Nature, 385, 409.
\bibitem{OI}
Okamoto I., 1999, MNRAS, 307, 253.
\bibitem{ODE}
O'Dell S.L., 1981, ApJ 243, L147.
\bibitem{PNICE}
Pelletier G., 1985, ConfŽrence de la SociŽtŽ Fran\c caise de Physique, Nice; and Pelletier G., Sol H., AssŽo E, 1988, Phys. Rev. A, 38, 2552.
\bibitem{PF}
Pelletier G., Ferreira J., Longaretti P-Y., 2004 (in preparation).
\bibitem{PP}
Pelletier G., Pudritz R.E., (1992), {\it ApJ.}, {\bf Vol. 394}, pp. 117 
\bibitem{PERO}
Pelletier G., Rolland J., 1989, A \& A, 224, 241.
\bibitem{PESO}
Pelletier G., Sol H., 1992, MNRAS, 254, 635.
\bibitem{PHGS}
Pelletier G., Henri G., Gialis D., Saug\'e L., 2004, International Symposium On Gamma Ray Astronomy, Heidelberg (in press); and SaugŽ L., Henri G., 2004, ApJ (submitted).
\bibitem{PS}
Phinney E.S., 1982, MNRAS 198, 1109.
\bibitem{PR}
Pudritz R., 2004, in "Accretion Disk, Jets and High Energy Phenomena in Astrophysics", Les Houches session 78 (EDP Sciences and Springer).
\bibitem{PN}
Pudritz R.\& Norman C., 1986, ApJ, 301, 571.
\bibitem{PU}
Punsly B., Coroniti F., 1990, ApJ, 350, 518.
\bibitem{PC}
Punsly B., Coroniti F., 1990, ApJ, 354, 583.
\bibitem{PL}
van Putten M., Levinson A., 2003, ApJ, 584, 937.
\bibitem{VP}
van Putten M., 2001, Phys. Rep., 345, 1.
\bibitem{RH}
Renaud N., Henri G., 1998, MNRAS, 300, 104.
\bibitem{R}
Rees M., 1967, MNRAS, 135, 345.
\bibitem{RM}
Rees M.J. \& Meszaros P., 1992, MNRAS, 258, 41.
\bibitem{SH}
Saug\'e L., Henri G., 2004, ApJ., 616, 136.
\bibitem{S}
Sauty C., Tsinganos K. \& Trussoni E., 1998, Ap\& SS.261.151S. 
\bibitem{SS}
Shakura N., Sunyaev R., 1973, A \& A, {\bf Vol. 24}, pp. 337.
\bibitem{SPA}
Sol H., Pelletier G., AssŽo E., 1989, MNRAS, 237, 411.
\bibitem{STRO}
Strominger A., 1998, JHEP, 02, 009.
\bibitem{SW}
Swenson R., 1987, MNRAS, 227, 403.
\bibitem{TP}
Tagger M., Pellat R., 1999 A\& A 349, 1003.
\bibitem{TPD}
Thorne K.S., Price R.H., McDonald D.A., 1986, "Black Hole: the Membrane Paradigm".
\bibitem{TB}
Tsinganos,ÊK.; Bogovalov,ÊS., 2002, MNRAS, 337, 553.
\bibitem{VT}
Varni\`ere P., Tagger M., 2002, A \& A, 394, 329.
\bibitem{VE}
Velikhov, E. (1959), {\it Sov. Phys. JETP}, {\bf Vol. 36}, pp. 1398.
\bibitem{VIET}
Vietri M., 2001, AIPC, 599, 416.
\bibitem{VL}
Vlahakis N., 2004, ApJ, 600, 324.
\bibitem{VK}
Vlahakis N., K\"onigl A., 2003, ApJ, 596, 1080.
\bibitem{VTT}
Vlahakis N., Tsinganos K., Sauty C., Trussoni E., 2000, MNRAS, 318, 417.
\bibitem{WAX}
Waxman E., 2000, ApJS, 127, 519.
\bibitem{WIL}
Wilson A.S., Young A.J., Shopbell P.J.: 2000, ApJ Lett. 544, L30.
\bibitem{ZN}
Zeldovich Ya., Novikov I., "Relativistic Astrophysics".

\end{thebibliography}
\end{document}